\documentclass[journal]{IEEEtran}
\usepackage{cite}
\usepackage[utf8]{inputenc}
\usepackage[T1]{fontenc}
\usepackage{graphicx}
\usepackage{amsmath,amsfonts,amssymb}
\usepackage{parskip}
\usepackage[table]{xcolor}
\usepackage{bm}
\usepackage{svg}
\usepackage{tikz}
\usepackage{tikz-3dplot}
\usepackage{soul}
\usepackage{pgfplots}
\usepackage{hyperref}
\usepackage{subcaption}
\usepackage{colortbl}
\usepackage{booktabs}   
\usepackage{multirow}   
\usepackage{threeparttable}
\usepackage{rotating}   
\usepackage{colortbl}   
\usepackage{siunitx}    
\usepackage{pgfplotstable} 
\pgfplotsset{compat=newest}
\usetikzlibrary{arrows.meta,decorations.markings,angles,quotes,3d,perspective}
\def\BibTeX{{\rm B\kern-.05em{\sc i\kern-.025em b}\kern-.08em
		T\kern-.1667em\lower.7ex\hbox{E}\kern-.125emX}}

\title{Understanding the Geometry of Faulted Power Systems under High Penetration of Inverter-Based Resources via Ellipse Fitting and Geometric Algebra}
\author{
\IEEEauthorblockN{Jorge Ventura, Jaroslav Hrdina, Aleš Návrat, Marek Stodola, Ahmad Eid, \\ 
Santiago Sanchez-Acevedo and Francisco G. Montoya}
}

\begin{document}
\maketitle

\begin{abstract}
Power systems with high penetration of inverter-based resources (IBR) present significant challenges for conventional protection schemes, with traditional distance protection methods failing to detect line-to-line faults during asymmetric conditions. This paper presents a methodology for electrical fault detection and classification using ellipse fitting and geometric algebra applied to voltage and current space curves. The approach characterizes electrical faults by fitting ellipses to voltage vector data, enabling fault detection with only a quarter-cycle. The method employs bivector components for line-to-ground fault classification, while ellipse parameters identify line-to-line and three-phase faults. The geometric representation preserves voltage or current curve shapes in three-dimensional space, overcoming Clarke transform limitations when zero-sequence components are present. Validation using simulations and laboratory experiments demonstrates accurate fault identification and magnitude estimation, providing enhanced power system protection capabilities.

\end{abstract}


\section{Introduction}
	In power systems, the identification and the classification of faults represent critical challenges for system protection and reliability. Modern electric power systems are integrating more renewable energy sources (RES) such as wind farms and photovoltaic parks. These RES are connected to the power system with power electronics inverters known as inverter-based-resources (IBR). The high penetration of IBRs presents new challenges to classic protection algorithms. The fault currents presented in systems with IBRs do not behave as the case of traditional power systems \cite{ConverterDistanceProtection1}.
    Thus,  robust methods for fault detection and classification are demanded. Electrical faults such as line-to-ground, line-to-line, and three-phase faults compromise system performance, causing failures in sensitive equipment and resulting in economic losses \cite{bollen2000understanding}. Traditional approaches based on RMS calculations and symmetric component transformations show significant limitations when dealing with noise, transients, or asymmetrical conditions. Moreover, symmetric component analysis requires phasor computation through Fourier coefficient extraction, and for non-sinusoidal systems, necessitates harmonic-by-harmonic decomposition, resulting in significant computational overhead. 
    
    Different authors have identified that the distance protection function tends to underperform in power systems with IBRs. The authors in \cite{ConverterDistanceProtection1} studied operation strategies of IBRs to inject fault current that mitigate distance protection misoperation, as recommended in three grid codes. Additionally, the work tested one more strategy that is not defined in the grid codes. The results highlighted that distance protection did not respond properly with current injection strategies defined in the grid codes.
 In addition, line-to-line faults (LLF) are not detected by traditional distance protection methods in power systems with IBRs \cite{ConverterDistanceProtection2}. The authors tested four different vendor distance protection devices in the case of an IBR with type-4 wind turbines \cite{ConverterDistanceProtection2}. The results show that distance protection fails to identify the fault in the case of LLF, caused by the lack of negative sequence current from the IBRs during the fault. This missed LLF is a result of the IBR control algorithm. Other works \cite{ConverterDistanceProtection3,ConverterDistanceProtection4,ConverterDistanceProtection5, Andrzej1} evaluate the performance of distance protection for two types of converter controllers. First control strategy is the grid forming converter \cite{ConverterDistanceProtection3,ConverterDistanceProtection4,ConverterDistanceProtection5} and the second strategy is grid following \cite{ConverterDistanceProtection3,ConverterDistanceProtection4,ConverterDistanceProtection5, Andrzej1}. It has been concluded that different control strategies of IBRs result in different distance protection performance. The challenge lies in the established interpretation of voltage and current dynamics during faults that under high IBR penetration still lead the protection system to make incorrect decisions.
 
Therefore, advanced analytical methods are required. Among the available techniques, geometric interpretations—specifically, ellipse fitting—offer an effective approach for evaluating voltage waveforms during fault conditions in polyphase systems \cite{Alam2015ANA} and for characterizing three-phase electrical networks \cite{9637952}. This approach allows for a precise representation of electrical faults, facilitating their classification and diagnosis.
Several studies have adopted the use of ellipses to model and analyze three-phase voltages. A common line of research relies on the Clarke transform \cite{clarke1938problems} for dimensional reduction, with numerous works \cite{ma2021evaluation,CamarilloPearanda2018CharacterizationOV,Alam2015CharacterizingVS,GarcaSnchez2017ApproachTF,Bagheri2018SpacePM} employing it to analyze voltage curve deviations in two dimensions. However, this approach has limitations. Specifically, the Clarke transform geometrically performs a rotation, but when the system becomes unbalanced (as occurs during asymmetrical faults) the zero component ($0$) remains, which is often overlooked. By analyzing only the $\alpha \beta 0$ components, these techniques effectively perform a projection rather than a true rotation, which distorts the actual 3D geometry of the fault trajectory. Although \cite{Ignatova2009SpaceVM} considers this third component, it treats it independently of the other two, failing to analyze the actual three-dimensional shape of the voltage vector and fragmenting the analysis.

Other approaches have explored alternative geometries. For instance, \cite{Alam2015ANA} utilizes a 3D polarization ellipse, but its analysis is based on spherical coordinate projections, which do not preserve the original curve shape in the same manner as a direct rotation. Another geometric analysis is proposed in \cite{9637952} by using synchronized Lissajous curves to monitor stability and identify anomalies. The technique leverages the capabilities of GPS-synchronized Waveform Measurement Units (WMUs) to construct graphical representations that facilitate the early detection of faults. However, this method requires both voltage and current data from at least two distinct WMUs, limiting its applicability to single-point measurements. 
Crucially, none of the aforementioned methods have been validated or tested in systems with high penetration of IBRs, where fault dynamics are significantly different and pose unique challenges to classical protection schemes.

Recent advances in Geometric Algebra (GA) applications to power systems, including coordinate transformation methods such as the Montoya Transform \cite{montoyaTransform} for unbalanced systems, demonstrate its versatility in electrical engineering contexts \cite{eid2022systematic,eid2024,bh}. While previous works such as \cite{Alam2020ClassificationAL, Li2019FastIM, CamarilloPearanda2018FaultCA} have established effective classification techniques for voltage sags based on ellipse parameters, our approach introduces a novel GA-based analysis that enhances both detection speed and classification accuracy for various types of faults. This is particularly relevant for asymmetric faults, which traditionally present challenges for existing methods, especially with the integration of IBRs. This paper combines robust ellipse fitting with GA to address the limitations of classic fault detection. GA provides a rigorous, yet intuitive approach for representing and manipulating conic sections, enabling the precise characterization of electrical faults in both time and space. Unlike conventional techniques, our method directly fits elliptical models to voltage trajectories in their native geometric form, improving robustness, accuracy, and computational efficiency for power systems with high penetration of IBRs. The methodology is validated using simulation studies and laboratory experiments demonstrating enhanced fault identification capabilities for modern power system protection.

The following are key contributions of our work:
\paragraph{A Novel Geometric Methodology} We propose a method that combines a true spatial rotation  with a robust ellipse fitting algorithm using Geometric Algebra. This preserves the geometric integrity of voltage and current trajectories, overcoming the distortions inherent in the misinterpretation of methods like the Clarke transform as mere projections rather than true rotations, especially during asymmetric faults with zero-sequence components.
\paragraph{Ultra-Fast Fault Detection} The proposed technique accurately detects and classifies all major fault types using only a quarter-cycle of waveform data, offering a significant speed advantage over conventional methods that typically require at least one half or full cycle.
\paragraph{Effective IBR Fault Classification} The framework explicitly addresses documented failures of traditional protection in IBR-dominated grids. It reliably identifies not only ground faults through bivector analysis but also challenging line-to-line faults—often missed by distance relays in IBR systems—by analyzing the ellipse orientation and parameters.
\paragraph{Comprehensive Validation} The method's performance and robustness are rigorously validated across a wide range of scenarios, including simulations with IBR-based grid models, and experiments with physical converters in a laboratory setup.

This paper is organised as follows: section \ref{Sec: Proposed Approach} presents a theoretical background on GA and ellipse fitting for conics. Section \ref{sec:method} describes the methodology for analysing faults with ellipses. Section \ref{sec:results} shows simulation and experimental results used to validate and demonstrate the performance of the method. Finally, conclusions are highlighted in Section \ref{sec:conclusions}.

\section{Theoretical Background}\label{Sec: Proposed Approach}

The geometric representation of a three-phase, three-wire system's signals in the time domain is constructed from the instantaneous values of phase voltages or line currents ($x_k(t)$ for $k \in {a,b,c}$) in vector form according to:
\begin{equation}
	\bm{x}(t) = x_a(t)\bm{\sigma}_1+x_b(t)\bm{\sigma}_2+x_c(t)\bm{\sigma}_3
\end{equation}
\noindent where $\bm{\sigma}=\{\bm{\sigma}_1,\bm{\sigma}_2,\bm{\sigma}_3\}$ is an orthonormal Euclidean basis. Under sinusoidal excitation, this vector $\bm{x}$ traces a closed, flat curve in space. Furthermore, for balanced systems, this curve is a circle, contained in the so-called \textit{Kirchhoff plane}. This plane has normal vector $\bm{k}=\bm{\sigma}_1+\bm{\sigma}_2+\bm{\sigma}_3$ (see \cite{eid2022systematic}) and is naturally represented in geometric algebra by a bivector $\bm{K}$ (see Appendix A). For a general (possibly unbalanced) case, the plane containing the trajectory can be represented using the outer product (see Appendix A) between two vectors at different time instants $t_1$ and $t_2$:
\begin{equation}
	\bm{B} = \bm{x}_{t_1} \wedge \bm{x}_{t_2}
    \label{eq:planeB}
\end{equation}

\begin{figure}[t]
    \centering
    \begin{tikzpicture}
        \begin{axis}[
            width=0.96\columnwidth,
            height=4cm,
            xlabel={Vector angle (rad)},
            ylabel={Relative Mean  Error (\%)},
            xmin=0, xmax=6.28,
            ymin=0, ymax=50,
            xtick={0, 1.57, 3.14, 4.71, 6.28},
            xticklabels={$0$, $\frac{\pi}{2}$, $\pi$, $\frac{3\pi}{2}$, $2\pi$},
            grid=both,
            grid style={dotted, gray!30},
            legend style={
                at={(0.98,0.98)},      
                anchor=north east,     
                legend columns=-1,      
                font=\scriptsize,    
                draw=none,          
                fill=none,            
                text opacity=1,        
                cells={anchor=west},   
                legend cell align=left, 
                legend image post style={xscale=0.5},
            },
            samples=200,            
            ]
            
            \addlegendimage{empty legend}
            \addlegendentry{\hspace{-20pt}\textbf{Std. Dev.}}
            
            
            \addplot[color=blue, line width=0.8pt, mark=none, line join=round] table [
                x index=0, 
                y index=1, 
                col sep=comma,
                skip first n=1    
            ] {data/error_bivector_analysis.csv};
            \addlegendentry{$0.1\%$}
            
            \addplot[color=red, line width=0.8pt, mark=none, line join=round] table [
                x index=0, 
                y index=2, 
                col sep=comma,
                skip first n=1    
            ] {data/error_bivector_analysis.csv};
            \addlegendentry{$1.0\%$}
            
            \addplot[color=orange, line width=0.8pt, mark=none, line join=round] table [
                x index=0, 
                y index=3,  
                col sep=comma,
                skip first n=1    
            ] {data/error_bivector_analysis.csv};
            \addlegendentry{$2.0\%$}
            
            \addplot[color=purple, line width=0.8pt, mark=none, line join=round] table [
                x index=0, 
                y index=4,  
                col sep=comma,
                skip first n=1    
            ] {data/error_bivector_analysis.csv};
            \addlegendentry{$5.0\%$}
            
            \addplot[color=green!60!black, line width=0.8pt, mark=none, line join=round] table [
                x index=0, 
                y index=5,  
                col sep=comma,
                skip first n=1    
            ] {data/error_bivector_analysis.csv};
            \addlegendentry{$10.0\%$}
            
        \end{axis}
    \end{tikzpicture}
    \caption{Relative mean error in bivector component estimation as a function of vector angle for different measurement noise levels (standard deviation from 0.1\% to 10.0\%).}
    \label{fig:bivector_error}
\end{figure}
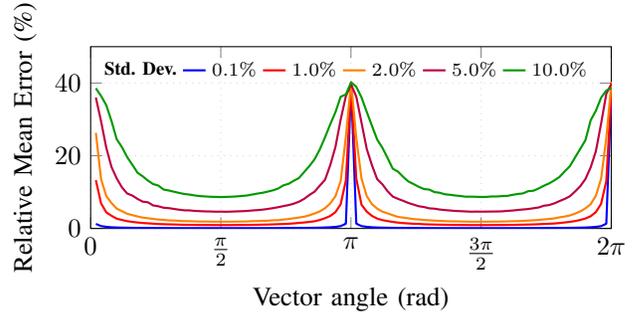

In theory, Eq. \eqref{eq:planeB} provides exact results regardless of time separation. However, due to the underlying geometric constraints, bivector accuracy becomes sensitive to the time separation of $\bm{x}(t_1)$ and $\bm{x}(t_2)$ in realistic scenarios with noise. 
Fig.\ref{fig:bivector_error} demonstrates how error in bivector component estimation is minimized when the angle between input vectors corresponds to approximately 25\% or 75\% of a cycle ($\frac{\pi}{2}$ or $\frac{3\pi}{2}$ radians), representing orthogonal vector orientations. Conversely, maximum errors occur when vectors approach parallelism at the beginning, middle, and end of the cycle.

The orientation of the plane containing the curve varies with system unbalance, providing a direct indicator of fault type. For line-to-ground and line-to-line-to-ground faults, the deviation of the plane from the nominal Kirchhoff plane is significant and characteristic. By analysing the normalized bivector components, these types of faults can be classified with high accuracy based solely on the orientation of their plane, as illustrated in Fig. \ref{fig:fault_classification}.

However, for line-to-line and three-phase faults, the trajectory plane remains largely coincident with the Kirchhoff plane. In these cases, the fault information can not captured by the direction of the bivector but is instead encoded in the shape of the voltage curve itself, which deforms from a circle into an ellipse. 
To accurately analyse this elliptical shape while preserving its geometric integrity, a true spatial rotation is required. This rotation aligns the plane represented by bivector ${\boldsymbol{B}}$ with a canonical reference plane, such as the $xy$ plane represented by the bivector $\bm{\sigma}_{12}$, without altering the intrinsic shape of the curve. This geometric approach aligns with coordinate transformation methodologies such as the Montoya Transform \cite{montoyaTransform}, which similarly leverages bivector representations and geometric rotors to identify and align planes containing voltage loci in unbalanced systems. This operation is performed using a geometric rotor $\bm{R}$, a specific operator from Geometric Algebra. The detailed construction of such a rotor and its associated algebraic properties are presented in Appendix A.
The rotation can be also applied to the space curve defined by vector $\bm{x}$ (or any general multivector) through the sandwich product:
    \begin{equation}
        \bm{x}_T = \bm{RxR}^{\dagger}
        \label{Apply_rotation}
    \end{equation}
\noindent resulting in the same space curve but defined with new coordinates by vector $\bm{x}_T$. This approach achieves dimensional reduction while preserving the original curve shape.

When a fault occurs, the curve usually adopts an elliptical shape, with both the plane orientation and the ellipse parameters providing critical information about the type of fault and the severity of the imbalance.
For precise determination of ellipse parameters from signal samples, we employ the Geometric Algebra for Conics (GAC) model developed in \cite{fit,GAC}. This GA defines an eight-dimensional geometric space $G_{5,3}$ where any conic in $\mathbb{R}^2$ can be represented as a vector using only 6 coordinates:

\begin{equation}\label{coninc_vector}
\bm{Q} = v_1\bm{\sigma}_0 + v_2\bm{\tilde\sigma}_0 + v_3\bm{\bar\sigma}_0 + v_4\bm{\sigma}_1 + v_5\bm{\sigma}_2 + v_6\bm{\sigma}_\infty
\end{equation}

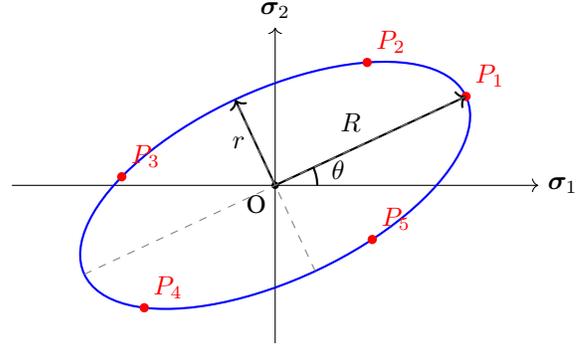
\begin{figure}[t]
    \centering

    \begin{tikzpicture}[scale=0.7]
        \def\R{4}        
        \def\r{1.8}      
        \def\ellipangle{25}   
        
        \draw[->] (-5,0) -- (5,0) node[right] {$\bm{\sigma}_1$};
        \draw[->] (0,-3) -- (0,3) node[above] {$\bm{\sigma}_2$};
        
        \draw[thick, blue] plot[domain=0:360, samples=100] 
            ({\R*cos(\x)*cos(\ellipangle) - \r*sin(\x)*sin(\ellipangle)},
             {\R*cos(\x)*sin(\ellipangle) + \r*sin(\x)*cos(\ellipangle)});
        
        \fill[black] (0,0) circle (2pt) node[below left] {O};
        
        \foreach \angle/\label in {0/P_1, 50/P_2, 130/P_3, 216/P_4, 288/P_5} {
            \pgfmathsetmacro{\px}{\R*cos(\angle)*cos(\ellipangle) - \r*sin(\angle)*sin(\ellipangle)}
            \pgfmathsetmacro{\py}{\R*cos(\angle)*sin(\ellipangle) + \r*sin(\angle)*cos(\ellipangle)}
            \fill[red] (\px,\py) circle (2.5pt);
            \node[red] at (\px,\py) [above right] {$\label$};
        }
        
        \pgfmathsetmacro{\Rx}{\R*cos(\ellipangle)}
        \pgfmathsetmacro{\Ry}{\R*sin(\ellipangle)}
        \draw[thick, black, ->] (0,0) -- (\Rx,\Ry) node[midway, above left] {$R$};
        
        \pgfmathsetmacro{\rx}{-\r*sin(\ellipangle)}
        \pgfmathsetmacro{\ry}{\r*cos(\ellipangle)}
        \draw[thick, black, ->] (0,0) -- (\rx,\ry) node[midway, left] {$r$};
        
        \draw[thick, black] (0.8,0) arc (0:\ellipangle:0.8);
        \node[black] at (1.2,0.3) {$\theta$};
        
        \draw[dashed, gray] (-\Rx,-\Ry) -- (\Rx,\Ry);
        \draw[dashed, gray] (-\rx,-\ry) -- (\rx,\ry);
    \end{tikzpicture}

    \caption{Ellipse (a type of conic section) uniquely determined by five given points.}
    \label{fig_exp:hyperbola}
\end{figure}

This formulation enables the construction of any conic from a minimum of five points, as illustrated in Fig.\ref{fig_exp:hyperbola}. From the vector coefficients $v_i$, we determine the conic type and derive its parameters. 

 The resulting ellipse $\bm{Q}$ is expressed as a vector according to  equation \eqref{coninc_vector}. The values of the semi-axes and the angle of inclination are derived from its coefficients. The angle $\theta$ represents the inclination of the major semi-axis relative to the principal direction of the space defined by $\bm{\sigma}_1$.

    \begin{equation}\label{eq:ellip_angle}
        \mathbf{\theta} = \frac{1}{2} \arctan \frac{v_3}{v_2}
    \end{equation}

    On the other hand, the major and minor semi-axes are calculated as follows:

    \begin{equation}\label{eq:ellip_semiaxes}
        a = \sqrt{\frac{\beta}{1-\alpha}};\quad b = \sqrt{\frac{\beta}{1+\alpha}}
    \end{equation}
    
    where $\alpha$ and $\beta$ are obtained such that
    
    \begin{equation}\label{eq:ellip_alpha&beta}
        \mathbf{\alpha} =\sqrt{v_2^2 + v_3^2};\quad \beta = -2v_6
    \end{equation}

Upon determining the elliptical parameters, comprehensive fault classification becomes feasible across all power system fault typologies. The ellipse orientation angle serves as a crucial discriminator for fault conditions that are indistinguishable through bivector coefficient analysis alone. 
The quantification of fault severity is subsequently derived from the minor semi-axis parameter.

\section{Methodology for detecting faults with ellipses}\label{sec:method}

    Whereas the Clarke transform is conventionally employed as a dimensional reduction technique in power system analysis, it introduces analytical inaccuracies when unbalance is present, as it distorts the three-dimensional trajectory through projection. The proposed rotational transformation, in contrast to conventional methods, facilitates the alignment of any arbitrary plane containing the vector ensemble with the $xy$ coordinate plane (bivector $\bm{\sigma}_{12}$ in GA), thereby achieving dimensionality reduction while preserving the geometric integrity of the signal locus.
    
    \subsection{Signal Mapping and Bivector extraction}

\begin{figure}[t]
        \centering
        \begin{tikzpicture}
            \begin{axis}[
                width=\columnwidth,
                height=0.5\columnwidth,
                xlabel={Time (s)},
                ylabel={Voltage (V p.u.)},
                grid=major,
                xmin=0.08, xmax=0.12,
                ymin=-1.2, ymax=1.2,
                legend style={at={(0.97,0.97)}, anchor=north east, font=\footnotesize, draw=black, fill=white, align=left},
                legend cell align={left},
                tick align=outside,
                xticklabel style={/pgf/number format/fixed},
                ]
                
                \addplot [thick, blue] table [x=t, y=va, col sep=comma] {data/fault_A_B.csv};
                \addplot [thick, red] table [x=t, y=vb, col sep=comma] {data/fault_A_B.csv};
                \addplot [thick, green!70!black] table [x=t, y=vc, col sep=comma] {data/fault_A_B.csv};
                
                \pgfmathsetmacro{\windowstart}{0.09}
                \pgfmathsetmacro{\windowend}{0.095}
                \pgfmathsetmacro{\windowwidth}{\windowend-\windowstart}
                
                \draw[fill=yellow, fill opacity=0.2, draw=orange, thick] 
                    (axis cs:\windowstart,-1.1) rectangle (axis cs:\windowend,1.1);
                
                \node[draw=none, fill=none, text=black, align=center, font=\scriptsize,
                      anchor=north west, inner sep=2pt] 
                    at (axis cs:\windowstart-0.0002,1) {Sliding\\Window};

                \draw[-{Stealth[length=6pt, width=4pt]}, orange, ultra thick] 
                    (axis cs:\windowstart+0.0005,0) -- (axis cs:\windowend-0.0005,0);

                \draw [dashed, red, thick] (axis cs:0.1,-1.2) -- (axis cs:0.1,1.2);
                \node[draw, fill=white, text=red, align=center, font=\small, 
                      anchor=north, inner sep=2pt] 
                    at (axis cs:0.1,-1.2) {Fault};
                
                \legend{Phase A, Phase B, Phase C}
            \end{axis}
        \end{tikzpicture}
        \caption{Visualization of normalized three-phase voltage signal with a A-B fault occurring at $t=0.1$s. The yellow rectangle represents a sliding window of 1/4 cycle for analysis.}
        \label{fig:sliding_window}
    \end{figure}
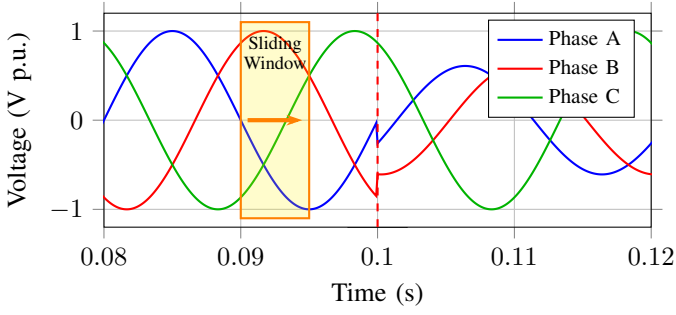

     The methodology begins by sampling voltage and/or current signals at a fixed rate $f_s$ using a 
     data acquisition system. To ensure precise disturbance classification, the raw signals undergo preprocessing, which includes anti-aliasing filtering and noise suppression, to maintain a signal-to-noise ratio (SNR) above 30 dB. For real-time waveform analysis, a quarter-cycle moving window is implemented. The rationale for this window duration, as substantiated in Section \ref{Sec: Proposed Approach}, is that it represents the most effective data span necessary for the optimal operation of the proposed algorithm. Nevertheless, a reduced window size may be employed, trading off precision for computational speed. Fig.\ref{fig:sliding_window} illustrates the three-phase voltage waveforms during an A-B line-to-line fault, overlaid with the corresponding analysis window.

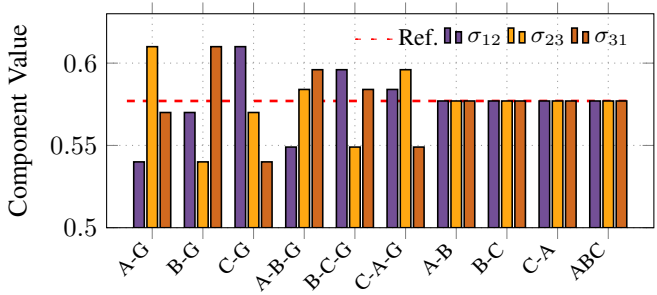
\begin{figure}[t]
    \centering
    \definecolor{darktangerine}{rgb}{1.0, 0.66, 0.07}
    \definecolor{cinnamon}{rgb}{0.82, 0.41, 0.12}
    \definecolor{darklavender}{rgb}{0.45, 0.31, 0.59}
    \begin{tikzpicture}
        \begin{axis}[
            width=\columnwidth,
            height=0.5\columnwidth,
            xlabel={},
            ylabel={Component Value},
            ymin=0.5, ymax=0.63,
            grid=both,
            minor grid style={dotted, gray!70},
            major grid style={dotted, gray!80},
            ybar=1pt,
            bar width=4pt,
            xtick={0,1,2,3,4,5,6,7,8,9,10},
            xticklabels={Ref,A-G, B-G, C-G, A-B-G, B-C-G, C-A-G, A-B, B-C, C-A, ABC},
            x tick label style={
                                rotate=45,               
                                anchor=east,             
                                font=\footnotesize,      
                                },
            legend style={
                            at={(0.98,0.98)},       
                            anchor=north east,      
                            legend columns=4,       
                            font=\small,            
                            draw=none,              
                            fill=none,              
                            text opacity=1,         
                            cells={anchor=west},    
                            legend image post style={scale=0.7}, 
                        },
            ]

            \addlegendimage{red, line legend, dashed}
            \addlegendentry{Ref.}
           
            \addplot[red, dashed, line width=1pt, sharp plot, update limits=false, forget plot] 
            coordinates {(0.5, 0.577) (10.5, 0.577)};

            \addplot[color=darklavender, fill=darklavender, draw=black, line width=0.6pt] coordinates {
                (1,0.54)
                (2,0.57)
                (3,0.61)
                (4,0.549)
                (5,0.596)
                (6,0.584)
                (7, 0.577) 
                (8, 0.577) 
                (9, 0.577) 
                (10, 0.577) 
            };
            \addlegendentry{$\sigma_{12}$}
            
            \addplot[color=darktangerine, fill=darktangerine, draw=black, line width=0.6pt] coordinates {
                (1,0.61)
                (2,0.54)
                (3,0.57)
                (4,0.584)
                (5,0.549)
                (6,0.596)
                (7, 0.577) 
                (8, 0.577) 
                (9, 0.577) 
                (10, 0.577) 
            };
            \addlegendentry{$\sigma_{23}$}
            
            \addplot[color=cinnamon, fill=cinnamon, draw=black, line width=0.6pt] coordinates {
                (1,0.57)
                (2,0.61)
                (3,0.54)
                (4,0.596)
                (5,0.584)
                (6,0.549)
                (7, 0.577) 
                (8, 0.577) 
                (9, 0.577) 
                (10, 0.577) 
            };
            \addlegendentry{$\sigma_{31}$}

        \end{axis}
    \end{tikzpicture}
    \caption{Classification of electrical faults based on bivector components}
    \label{fig:fault_classification}
\end{figure}

    The collection of $n$ data samples, indexed by $i$ and contained within the specified window length, is transformed into three-dimensional (3D) vectors. These vectors represent discrete points that generate the curve,
    \begin{equation}        
       \{{\bm{x}}_i\}_{i=1}^n,\ {\bm{x}}_i \in \mathbb{R}^3 \implies {\bm{x}}_i = x_{i,A}\bm{\sigma}_1 + x_{i,B}\bm{\sigma}_2 + x_{i,C}\bm{\sigma}_3
    \end{equation} 
    For each window frame, the vectors representing the first and last data points are utilized to construct the bivector that encapsulates the trajectory in the considered plane
    \begin{equation}        
        \bm{B} = {\bm{x}}_1 \wedge {\bm{x}}_n 
    \end{equation}    
    Bivector $\bm{B}$ can be normalized, resulting in $\hat{\bm{B}}=\bm{B}/\|\bm{B}\|$. Thus, a preliminary fault classification can be conducted based on the quantitative deviation of bivector coefficients $\hat{B}_i$ from the nominal value of the normalized Kirchhoff plane $\hat{\bm{K}}$ ($\hat{K}_i=1/\sqrt{3})$). Fig.~\ref{fig:fault_classification} illustrates the characteristic bivector patterns for common fault types, simulated using the Kundur two-area system model~\cite{kundur2007power}. The results correspond to steady-state conditions during the fault, after transient oscillations have decayed and the system reaches a permanent regime with the fault persisting.

    This initial analysis facilitates the identification of ground fault conditions exclusively. Subsequently, the rotor $\bm{R}$ that aligns plane $\hat{\bm{B}}$ and $\bm{\sigma}_{12}$ is computed in accordance with expression \eqref{eq:Rotations_components}.  Finally, the computed rotor is applied to the vector ensemble. 
    The time dependent vector is consequently represented in a two-dimensional subspace defined by the orthonormal basis $\{\bm{\sigma}_1,\bm{\sigma}_2\}$. This vector rotates about the origin at the fundamental frequency of the original three-phase system. Consequently, the vector at each discrete time $i$ are characterized by two scalar components using \eqref{Apply_rotation}, with their transformed coefficients expressed as
    
    \begin{equation}
        \bm{{x}_{i,T}} = \bm{Rx}_i\bm{R}^{\dagger}=x_{i,T_1} \bm{\sigma}_1 + x_{i,T_2}\bm{\sigma}_2
    \end{equation}

    \begin{figure}[]
    \centering
    \includegraphics[width=0.8\columnwidth]{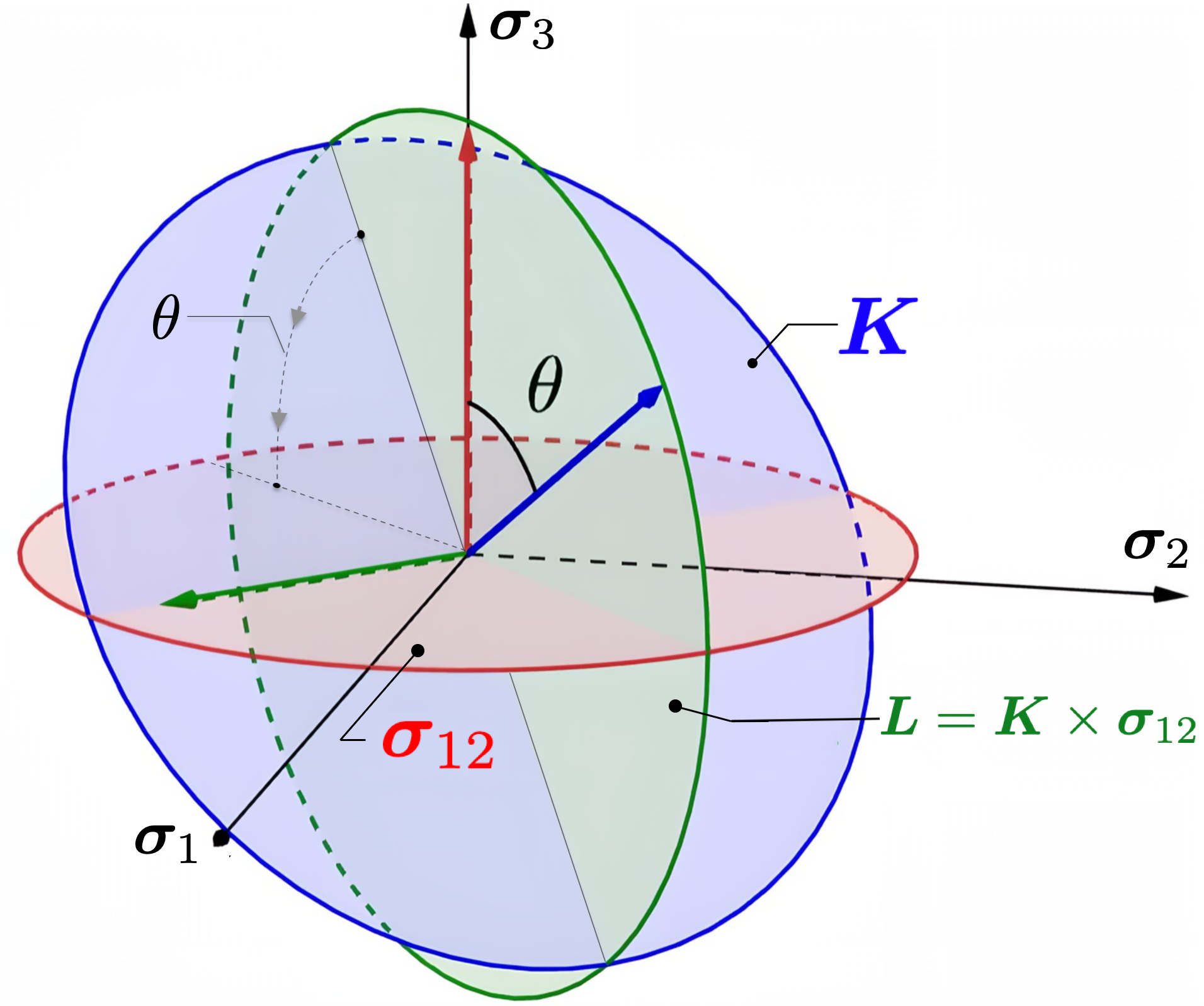}
    \caption{Rotation aligning the Kirchoff plane (represented by bivector $\bm{K}$) with the horizontal plane (represented by bivector $\bm{\sigma}_{12}$). The operation $\bm{K} \times \bm{\sigma}_{12}$ results in a bivector $\bm{L}$ that represents the plane of rotation where $\theta$ is the angle of rotation.}
    \label{fig:plane_rotation}
\end{figure}

Fig.\ref{fig:plane_rotation} provides a graphical representation of this spatial transformation process, illustrating the geometric relationships between the constituent planes involved in the process. Note that this operation rotates the entire plane as a rigid body, which inherently rotates any trajectory or curve lying on that plane.

    \subsection{GAC-Based Fitting Algorithm}

    With the set of samples in $\mathbb{R}^2$, we aim to identify the ellipse that accurately represents the complete state of the system for that  window frame. Ellipse fitting, based on the method proposed in \cite{GAC}, facilitates this process with significant precision by utilizing information from only a quarter of the ellipse. This approach offers an advantage over most studies that employ moving windows \cite{Ignatova2009SpaceVM,Alam2020ClassificationAL,Alam2015ANA,GarcaSnchez2017ApproachTF,Alam2015CharacterizingVS} spanning one cycle. Furthermore, the method can be configured to fit only ellipses centered at the origin, thereby enhancing precision by minimizing errors. The following are the steps taken in the adjustment algorithm using GAC \cite{Stodola2024}, incorporating the constraint of ellipses centered at the origin:
   
    \begin{enumerate}
        \item  Determine the matrix $B$ which represents the bilinear form associated with the inner product within the algebra $G^{5,3}$. 
        \begin{align*} 
            \label{eq:identities}
            B=
            \begin{pmatrix}
            0 & 0 & -1_{3 \times 3} \\ 0 & 1_{2 \times 2} & 0 \\ -1_{3 \times 3} & 0 & 0
            \end{pmatrix},
        \end{align*}
        \item Form the data matrix $D$ of size 8 x N, where $N$ is the column size and each column contains the 2D i-th point of the dataset in the form of a GAC vector $\bm{x}$.
        \begin{equation*}
            \bm{x} =\bm{\sigma}_0+x_1\bm{\sigma}_1+x_2\bm{\sigma}_2 + \frac{1}{2}\bm{x}^2 \bm{\sigma}_{\infty} + {1 \over 2}(x_1^2 - x_2^2) \tilde{\bm{\sigma}}_\infty + x_1x_2 \bar{\bm{\sigma}}_\infty,
        \end{equation*}
        \item Computation of the symmetric matrix P for the objective function 
        \begin{equation*}
            P = \frac{1}{N} BDD^TB
        \end{equation*}
        \item For an ellipse centred at the origin, set $v_4=v_5=0$ and define the reduced matrix $P_r$ as the principal submatrix of $P$ corresponding to the active parameters $\{v_1, v_2, v_3, v_6\}$.\\

        \item Apply the constraint matrix C to ensure that only ellipses are fitted through the restriction $v_3^2 - v_2^2-v_1^2 = 1$, which is equivalent to $4AC - B^2 =1 $ in standard form.
        \begin{equation*}
            C=
            \begin{pmatrix}
                -1 & 0 & 0 & 0 \\
                0 & -1 & 0 & 0 \\
                0 & 0 & 1 & 0 \\
                0 & 0 & 0 & 0 \\
            \end{pmatrix}
        \end{equation*}

        \item Solve the generalized eigenvalue problem $P_rv_i = \lambda_iC\bm{\mathrm{v}_i}$ and select the eigenvector $\bm{\mathrm{v}}^*$ corresponding to the smallest non-negative eigenvalue $\lambda^* = min\{\lambda_i:\lambda_i\geq 0 \}$.\\
        
        \item Construct vector $\bm{Q}$ and normalise it using the first coefficient from equation \eqref{coninc_vector}.
        \begin{equation*}
            \bm{Q} = {\bm{\mathrm{{v}}}}{v_1}^{-1}
        \end{equation*}
    \end{enumerate}

    Once the ellipse vector is obtained, its parameters (angle of inclination and semi-axes) are derived based on the equations \eqref{eq:ellip_angle}, \eqref{eq:ellip_semiaxes}, and \eqref{eq:ellip_alpha&beta}. 

\begin{figure}[]
    \centering

    \begin{tikzpicture}[scale=3]
        \draw[thick, ->] (-1.3,0) -- (1.3,0) node[right] {$\bm{\sigma_1}$};
        \draw[thick, ->] (0,-0.3) -- (0,1.3) node[above] {$\bm{\sigma_2}$};
        
        \fill[blue!30, opacity=0.7] (0,0) -- (1,0) 
            arc[start angle=0, end angle=15, radius=1] -- cycle;
        
        \fill[red!30, opacity=0.7] (0,0) -- ({cos(15)},{sin(15)}) 
            arc[start angle=15, end angle=75, radius=1] -- cycle;
        
        \fill[green!30, opacity=0.7] (0,0) -- ({cos(75)},{sin(75)}) 
            arc[start angle=75, end angle=135, radius=1] -- cycle;
        
        \fill[blue!30, opacity=0.7] (0,0) -- ({cos(135)},{sin(135)}) 
            arc[start angle=135, end angle=180, radius=1] -- cycle;
        
        \draw[thick, black] (1,0) arc[start angle=0, end angle=180, radius=1];
        \draw[thick, black] (-1,0) -- (1,0);
        
        \draw[dashed, thick] (0,0) -- ({1.05*cos(15)},{1.05*sin(15)});
        \node[font=\normalsize] at ({1.15*cos(15)},{1.15*sin(15)}) {$\frac{\pi}{12}$};
        \draw[dashed, thick] (0,0) -- ({1.05*cos(75)},{1.05*sin(75)});
        \node[font=\normalsize] at ({1.15*cos(75)},{1.15*sin(75)}) {$\frac{5\pi}{12}$};
        \draw[dashed, thick] (0,0) -- ({1.05*cos(135)},{1.05*sin(135)});
        \node[font=\normalsize] at ({1.15*cos(135)},{1.15*sin(135)}) {$\frac{3\pi}{4}$};
        
        \draw[thick] ({0.95*cos(45)},{0.95*sin(45)}) -- ({1.1*cos(45)},{1.1*sin(45)});
        \node[font=\large] at ({1.3*cos(45)},{1.3*sin(45)}) {$\bm{\frac{\pi}{4}}$};
        
        \draw[thick] ({0.95*cos(105)},{0.95*sin(105)}) -- ({1.1*cos(105)},{1.1*sin(105)});
        \node[font=\large] at ({1.3*cos(105)},{1.3*sin(105)}) {$\bm{\frac{7\pi}{12}}$};
        
        \draw[thick] ({0.95*cos(165)},{0.95*sin(165)}) -- ({1.1*cos(165)},{1.1*sin(165)});
        \node[font=\large] at ({1.3*cos(165)},{1.3*sin(165)}) {$\bm{\frac{11\pi}{12}}$};
        
        \node[font=\footnotesize, font=\bfseries] at ({0.7*cos(45)},{0.7*sin(45)}) {A-B};
        
        \node[font=\footnotesize, font=\bfseries] at ({0.7*cos(105)},{0.7*sin(105)}) {A-C};
        
        \node[font=\footnotesize, font=\bfseries] at ({0.7*cos(165)},{0.7*sin(165)}) {B-C};

    \end{tikzpicture}
    \caption{Classification of line-to-line faults based on the inclination angle of the ellipse}
    \label{fig:fault_classification_circle}
\end{figure}
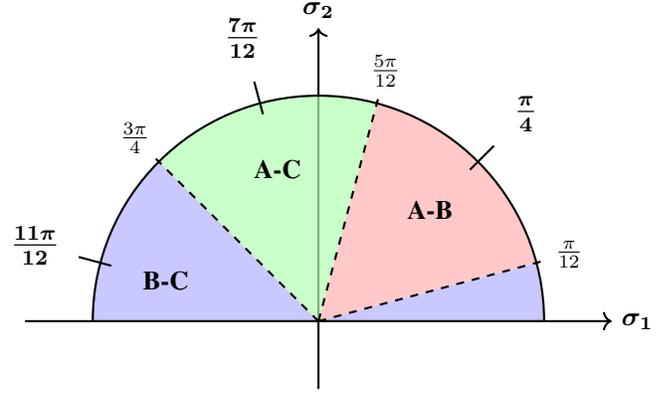

    For line-to-line asymmetric faults, the orientation of the ellipse indicates which phases are responsible. In the case of an ideal fault, this angle assumes fixed values, as shown in Table \ref{tab:fault_parameters_comparison}. However, for real faults, deviations from these values may occur, highlighting the need to define the range of values for each type of fault. For this work, these ranges have been established according to the sectorial representation in Fig.\ref{fig:fault_classification_circle}.
    
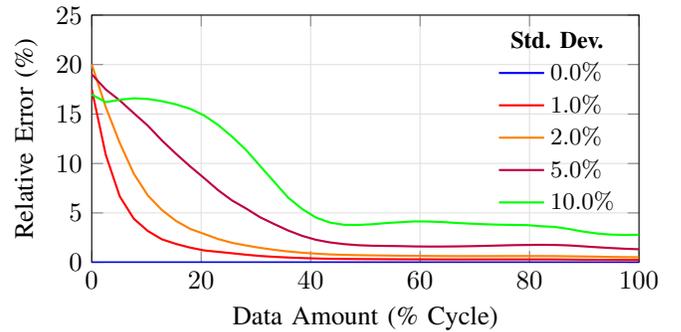
\begin{figure}[]
    \centering
    \begin{tikzpicture}
        \begin{axis}[
            width=\columnwidth,
            height=0.55\columnwidth,
            xlabel={Data Amount (\% Cycle)},
            ylabel={Relative Error (\%)},
            xmin=0, xmax=100,
            ymin=0, ymax=25,
            ytick={0, 5, 10, 15, 20, 25},
            yticklabel style={/pgf/number format/fixed},
            grid=both,
            minor grid style={gray!15},
            major grid style={gray!25},
            legend style={
                at={(0.98,0.98)},
                anchor=north east,
                legend columns=1,
                font=\small,
                draw=none,
                fill=none,
                text opacity=1,
                cells={anchor=west},
                legend cell align=left
            },
            y filter/.code={\pgfmathparse{#1*100}\pgfmathresult}
        ]
        
        \addlegendimage{empty legend}
        \addlegendentry{\hspace{-15pt}\textbf{Std. Dev.}}
        
        \addplot[thick, blue] table[x=porcentaje, y=sigma0, col sep=comma] {data/error_ajuste_elipse.csv};
        \addlegendentry{$0.0\%$}
        
        \addplot[thick, red] table[x=porcentaje, y=sigma0.01, col sep=comma] {data/error_ajuste_elipse.csv};
        \addlegendentry{$1.0\%$}
        
        \addplot[thick, orange] table[x=porcentaje, y=sigma0.02, col sep=comma] {data/error_ajuste_elipse.csv};
        \addlegendentry{$2.0\%$}
        
        \addplot[thick, purple] table[x=porcentaje, y=sigma0.05, col sep=comma] {data/error_ajuste_elipse.csv};
        \addlegendentry{$5.0\%$}
        
        \addplot[thick, green] table[x=porcentaje, y=sigma0.1, col sep=comma] {data/error_ajuste_elipse.csv};
        \addlegendentry{$10.0\%$}        
        \end{axis}
    \end{tikzpicture}
          \caption{Relative error in ellipse parameter estimation as a function of data availability for varying noise levels.}
    \label{fig:error_ellip_fitting}
\end{figure}

     Fig.\ref{fig:error_ellip_fitting} presents a quantitative analysis of fitting error magnitudes as a function of data window size, parametrized by various noise levels. The error metric exhibits an inverse relationship with the temporal span of the data window; notably, for a quarter-cycle window (25\%) operating under moderate noise conditions—achievable through appropriate filtering techniques—the algorithm demonstrates exceptional accuracy with fitting error magnitudes consistently below 1\%.

\begin{table}[]
\centering
\caption{Ellipse parameters and bivector coefficients for different fault types and severity levels (p.u)}
\resizebox{\columnwidth}{!}{%
\begin{tabular}{cccccccc}
\toprule
\textbf{Fault} & \textbf{Severity} & $\pmb{\sigma}_{12}$ & $\pmb{\sigma}_{23}$ & $\pmb{\sigma}_{31}$ & \textbf{$R$ Maj. axis} & \textbf{$r$ Min. axis} & \textbf{Angle (rad)} \\
\midrule
\multirow{4}{*}{A-G}
& 0.1 & 0.5560 & 0.6178 & 0.5560 & 1.2247 & 1.1446 & 1.2862 \\
    & 0.4 & 0.4472 & 0.7746 & 0.4472 & 1.2247 & 0.9129 & 1.1957 \\
    & 0.6 & 0.3651 & 0.8539 & 0.3651 & 1.2247 & 0.8165 & 1.1071 \\
    & 0.9 & 0.0990 & 0.9902 & 0.0990 & 1.2247 & 0.7141 & 0.8801 \\
\midrule
\multirow{4}{*}{B-G}
& 0.1 & 0.5560 & 0.5560 & 0.6178 & 1.2247 & 1.1446 & 0.2846 \\
    & 0.4 & 0.4472 & 0.4472 & 0.7746 & 1.2247 & 0.9129 & 0.3751 \\
    & 0.6 & 0.3651 & 0.3651 & 0.8539 & 1.2247 & 0.8165 & 0.4636 \\
    & 0.9 & 0.0990 & 0.0990 & 0.9902 & 1.2247 & 0.7141 & 0.6907 \\
\midrule
\multirow{4}{*}{C-G}
& 0.1 & 0.6178 & 0.5560 & 0.5560 & 1.2247 & 1.1446 & 2.3562 \\
    & 0.4 & 0.7746 & 0.4472 & 0.4472 & 1.2247 & 0.9129 & 2.3562 \\
    & 0.6 & 0.8539 & 0.3651 & 0.3651 & 1.2247 & 0.8165 & 2.3562 \\
    & 0.9 & 0.9902 & 0.0990 & 0.0990 & 1.2247 & 0.7141 & 2.3562 \\
\midrule
\multirow{4}{*}{A-B-G}
& 0.1 & 0.5369 & 0.5966 & 0.5966 & 1.1853 & 1.1023 & 0.7854 \\
    & 0.4 & 0.3780 & 0.6547 & 0.6547 & 1.0801 & 0.7559 & 0.7854 \\
    & 0.6 & 0.2774 & 0.6831 & 0.6831 & 1.0408 & 0.4915 & 0.7854 \\
    & 0.9 & 0.0705 & 0.7054 & 0.7054 & 1.0025 & 0.1225 & 0.7854 \\
\midrule
\multirow{4}{*}{B-C-G}
& 0.1 & 0.5965 & 0.5369 & 0.5965 & 1.1853 & 1.1023 & 2.9015 \\
    & 0.4 & 0.6509 & 0.3906 & 0.6509 & 1.0863 & 0.7348 & 2.9735 \\
    & 0.6 & 0.6804 & 0.2722 & 0.6804 & 1.0392 & 0.4899 & 3.0268 \\
    & 0.9 & 0.7053 & 0.0705 & 0.7053 & 1.0025 & 0.1225 & 3.1123 \\
\midrule
\multirow{4}{*}{C-A-G}
& 0.1 & 0.5966 & 0.5966 & 0.5369 & 1.1853 & 1.1023 & 1.8109 \\
    & 0.4 & 0.6547 & 0.6547 & 0.3780 & 1.0801 & 0.7559 & 1.7701 \\
    & 0.6 & 0.6831 & 0.6831 & 0.2774 & 1.0408 & 0.4915 & 1.7391 \\
    & 0.9 & 0.7054 & 0.7054 & 0.0705 & 1.0025 & 0.1225 & 1.6001 \\
\midrule
\multirow{4}{*}{A-B}
& 0.1 & 0.5774 & 0.5774 & 0.5774 & 1.2247 & 1.1023 & 0.7854 \\
    & 0.4 & 0.5774 & 0.5774 & 0.5774 & 1.2247 & 0.7348 & 0.7854 \\
    & 0.6 & 0.5774 & 0.5774 & 0.5774 & 1.2247 & 0.4899 & 0.7854 \\
    & 0.9 & 0.5774 & 0.5774 & 0.5774 & 1.2247 & 0.1225 & 0.7854 \\
\midrule
\multirow{4}{*}{B-C}
& 0.1 & 0.5774 & 0.5774 & 0.5774 & 1.2247 & 1.1023 & 2.8798 \\
    & 0.4 & 0.5774 & 0.5774 & 0.5774 & 1.2247 & 0.7348 & 2.8798 \\
    & 0.6 & 0.5774 & 0.5774 & 0.5774 & 1.2247 & 0.4899 & 2.8798 \\
    & 0.9 & 0.5774 & 0.5774 & 0.5774 & 1.2247 & 0.1225 & 2.8798 \\
\midrule
\multirow{4}{*}{C-A}
& 0.1 & 0.5774 & 0.5774 & 0.5774 & 1.2247 & 1.1023 & 1.8326 \\
    & 0.4 & 0.5774 & 0.5774 & 0.5774 & 1.2247 & 0.7348 & 1.8326 \\
    & 0.6 & 0.5774 & 0.5774 & 0.5774 & 1.2247 & 0.4899 & 1.8326 \\
    & 0.9 & 0.5774 & 0.5774 & 0.5774 & 1.2247 & 0.1225 & 1.8326 \\
\midrule
\multirow{4}{*}{A-B-C}
& 0.1 & 0.5774 & 0.5774 & 0.5774 & 1.1023 & 1.1023 & 0 \\
    & 0.4 & 0.5774 & 0.5774 & 0.5774 & 0.7348 & 0.7348 & 0 \\
    & 0.6 & 0.5774 & 0.5774 & 0.5774 & 0.4899 & 0.4899 & 0 \\
    & 0.9 & 0.5774 & 0.5774 & 0.5774 & 0.1225 & 0.1225 & 0 \\
\bottomrule
\end{tabular}}
\label{tab:fault_parameters_comparison}
\end{table}

\subsection{Fault Magnitude Estimation}

Disturbance magnitude quantification constitutes a critical parameter in power system imbalance analysis. This metric enables the application of analytical models, such as those developed in \cite{CamarilloPearanda2018CharacterizationOV}, for fault impedance calculation, fault location estimation, and voltage event classification as either sag or swell phenomena. The semi-axes variation provides a precise estimation methodology due to the established linear correlation between fault intensity and elliptical minor semi-axis magnitude. Table \ref{tab:semiaxes_variations} presents the boundary values of elliptical parameters corresponding to each fault classification category.

\begin{table}[]
\renewcommand{\arraystretch}{1.3}
\caption{Semi-Axis Parameters for Various Fault Conditions. The parameter $A$ is the nominal phase RMS value.}
\label{tab:ellipse_parameters}
\centering
\begin{tabular}{l cccc}
\toprule
\multirow{2}{*}{\textbf{Fault Type}} & \multicolumn{2}{c}{\textbf{Major Semi-Axis}} & \multicolumn{2}{c}{\textbf{Minor Semi-Axis}} \\
\cmidrule(lr){2-3} \cmidrule(lr){4-5}
& $R_{\max}$ & $R_{\min}$ & $r_{\max}$ & $r_{\min}$ \\
\midrule
L - G & $A\sqrt{3}$ & $A\sqrt{3}$ & $A\sqrt{3}$ & $A$ \\
L - L - G & $A\sqrt{3}$ & $A\sqrt{2}$ & $A\sqrt{3}$ & $0$ \\
L - L & $A\sqrt{3}$ & $A\sqrt{3}$ & $A\sqrt{3}$ & $0$ \\
L - L - L & $A\sqrt{3}$ & $0$ & $A\sqrt{3}$ & $0$ \\
\bottomrule
\end{tabular}
\label{tab:semiaxes_variations}
\end{table}

Under balanced system conditions, the fitted ellipse degenerates to a circular locus with equivalent semi-axes of magnitude $A\sqrt{3}$, where $A$ represents the phase voltage RMS value. This geometric property aligns with the theoretical values of $\alpha$ and $\beta$ components in the Clarke domain for balanced sinusoidal networks. Under fault conditions, the semi-axes dimensions decrease proportionally with imbalance severity to their respective minimum values. Non-linear characteristic behavior is observed exclusively in the minor semi-axis during line-to-ground faults and in the major semi-axis during two-line-to-ground faults. For computational efficiency, linear approximation has been implemented with a resulting mean absolute error of 7.13\%.

The bivector components provide a robust method for identifying all line-to-ground and two-line-to-ground faults at any disturbance level. However, for line-to-line faults where the plane of the curve remains constant regardless of fault severity, the ellipse inclination angle serves as the key differentiating parameter. This angle remains consistent across different unbalance severities for a given fault type, providing a reliable classification parameter. The fault severity can be accurately quantified using the minor semi-axis of the ellipse.

\section{Results and Discussion} \label{sec:results}

This section presents the results of fault detection and identification with GA for a simulated distribution grid with two grid-forming converters (GFM) and one grid-following converter (GFL). Additionally, the laboratory experimental validation is presented by applying asymmetrical faults between a grid equivalent and a 60 kVA converter.

\subsection{Distribution grid with inverter based resources}

\begin{figure}[]
    \centering
    \includegraphics[width=0.99\columnwidth]{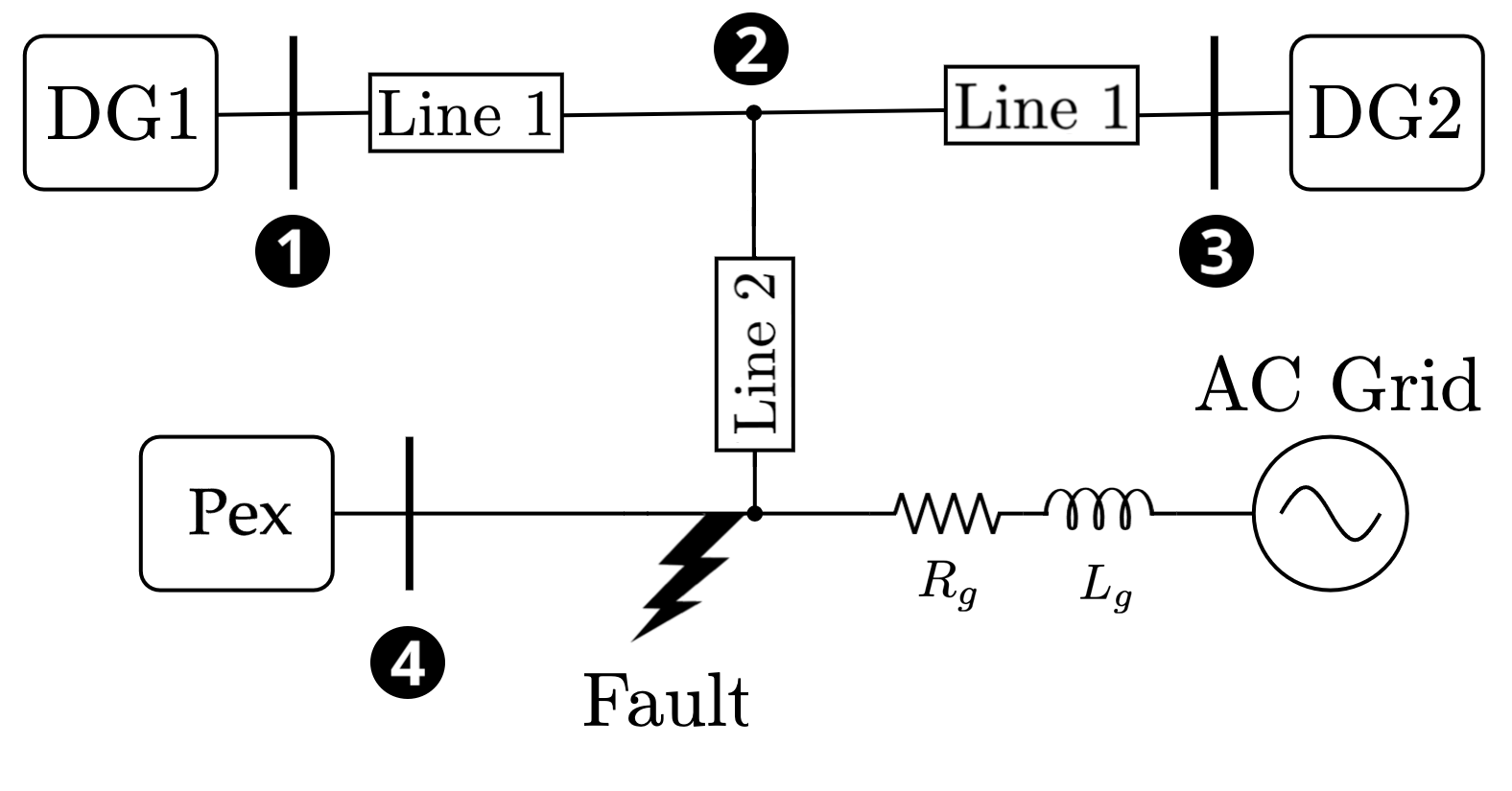}
    \caption{Distribution grid with two GFM and one GFL based resources.}
    \label{fig:grid}
\end{figure}

The distribution grid used in the simulation is based on the grid described in {\cite{Sanchez2015Stability}}. The AC grid comprises one grid equivalent, two distributed generators with GFM, and one GFL converter that connects renewable energy sources from a DC link to the AC grid. An illustration of the grid is provided in Fig. \ref{fig:grid}. The grid equivalent is characterized by an ideal source along with the grid equivalent resistance, denoted as $R_g$, and the grid equivalent inductance, represented as $L_g$. The two GFM converters are referred to as DG1 and DG2, while the GFL converter used for power exchange from the DC grid is designated as Pex. For the sake of simplicity, the DC source side of the converters use an ideal DC source.

The simulation was carried-out using Simulink software, with ideal switches for the converters and where all simulations have a sampling frequency of $10\ kHz$. The AC voltage and current variables for each converter were recorded. Subsequently, the proposed method was applied to extract the parameters defining these variables. 

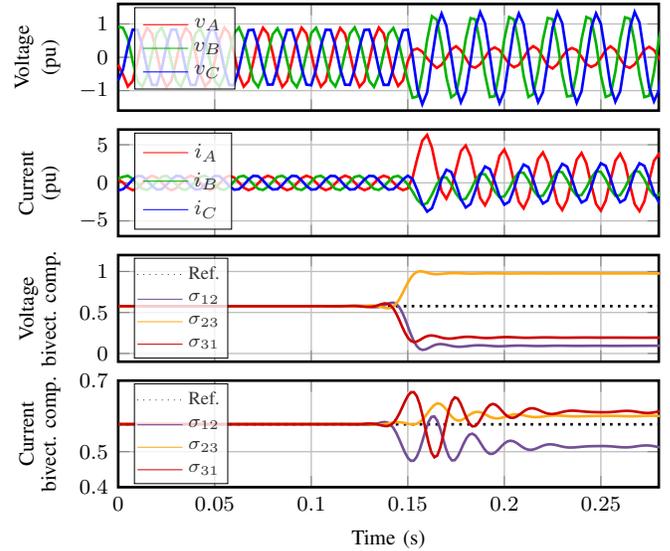
\begin{figure}[]
    \centering
    \definecolor{darktangerine}{rgb}{1.0, 0.66, 0.07}
    \definecolor{bostonuniversityred}{rgb}{0.8, 0.0, 0.0}
    \definecolor{darklavender}{rgb}{0.45, 0.31, 0.59}
    \begin{tikzpicture}
        \begin{axis}[
            width=0.99\columnwidth,
            height=3cm,
            xmin=0, xmax=0.28,
            ymin=-1.6, ymax=1.6,
            ylabel={\shortstack{Voltage\\(pu)}},
            ylabel style={font=\footnotesize},
            grid=major,
            grid style={line width=0.5pt, gray!50},
            line width=1pt,
            xticklabel=\empty,
            xtick={0,0.05,0.1,0.15,0.2,0.25,0.3,0.35,0.4,0.45,0.5},
            ytick={-1,0,1},
            tick label style={font=\footnotesize},
            legend pos=north west,
            legend style={font=\footnotesize, fill=white, fill opacity=0.8, line width=0.3pt},
            name=voltage_plot,
            enlarge x limits=false,
            enlarge y limits=false,
        ]
            \addplot[red, solid] table[x=t, y=va_DG1, col sep=comma] {data/Falla_AG_convert.csv};
            \addplot[green!70!black, solid] table[x=t, y=vb_DG1, col sep=comma] {data/Falla_AG_convert.csv};
            \addplot[blue, solid] table[x=t, y=vc_DG1, col sep=comma]{data/Falla_AG_convert.csv};
            
            \legend{$v_A$, $v_B$, $v_C$}
        \end{axis}
        \begin{axis}[
            width=0.99\columnwidth,
            height=3cm,
            xmin=0, xmax=0.28,
            ymin=-7, ymax=7,
            xlabel style={font=\footnotesize},
            ylabel={\shortstack{Current\\(pu)}},
            ylabel style={font=\footnotesize},
            grid=major,
            grid style={line width=0.5pt, gray!50},
            line width=1pt,
            at={(voltage_plot.below south west)},
            anchor=north west,
            yshift=0cm,
            xticklabel=\empty,
            xtick={0,0.05,0.1,0.15,0.2,0.25,0.3,0.35,0.4,0.45,0.5},    
            ytick={-5,0,5},
            tick label style={font=\footnotesize},
            legend pos=north west,
            legend style={font=\footnotesize, fill=white, fill opacity=0.8, line width=0.3pt},
            name=current_plot,
            enlarge x limits=false,
            enlarge y limits=false,
        ]
            \addplot[red, solid] table[x=t, y=ia_DG1, col sep=comma] {data/Falla_AG_convert.csv};
            \addplot[green!70!black, solid] table[x=t, y=ib_DG1, col sep=comma] {data/Falla_AG_convert.csv};
            \addplot[blue, solid] table[x=t, y=ic_DG1, col sep=comma] {data/Falla_AG_convert.csv};
            
            \legend{$i_A$, $i_B$, $i_C$}
        \end{axis}
        \begin{axis}[
            width=0.99\columnwidth,
            height=3cm,
            xmin=0, xmax=0.28,
            ymin=-0.1, ymax=1.2,
            ylabel={Voltage\\bivect. comp.},
            ylabel style={font=\footnotesize, align=center},
            grid=major,
            grid style={line width=0.5pt, gray!50},
            line width=1pt,
            at={(current_plot.below south west)},
            anchor=north west,
            yshift=0cm,
            xticklabel=\empty,
            xtick={0,0.05,0.1,0.15,0.2,0.25,0.3,0.35,0.4,0.45,0.5},
            ytick={0,0.5,1},
            tick label style={font=\footnotesize},
            legend pos=north west,
            legend style={font=\scriptsize, fill=white, fill opacity=0.8, line width=0.3pt,inner sep=1pt,  
    outer sep=0pt},
            name=voltage_biv_plot,
            enlarge x limits=false,
            enlarge y limits=false,
            ]
            \addplot[black, dotted] coordinates {(0, 0.577) (0.6, 0.577)};
            \addplot[darklavender, solid] table[x=t, y=s1_v, col sep=comma] {data/Falla_AG_bivec_convert.csv};
            \addplot[darktangerine, solid] table[x=t, y=s2_v, col sep=comma] {data/Falla_AG_bivec_convert.csv};
            \addplot[bostonuniversityred, solid] table[x=t, y=s3_v, col sep=comma] {data/Falla_AG_bivec_convert.csv};
            
            \legend{Ref.,$\sigma_{12}$, $\sigma_{23}$, $\sigma_{31}$}
        \end{axis}
        \begin{axis}[
            width=0.99\columnwidth,
            height=3cm,
            xmin=0, xmax=0.28,
            ymin=0.4, ymax=0.7,
            xlabel={Time (s)},
            xlabel style={font=\footnotesize},
            ylabel={Current\\bivect. comp.},
            ylabel style={font=\footnotesize, align=center},
            grid=major,
            grid style={line width=0.5pt, gray!50},
            line width=1pt,
            at={(voltage_biv_plot.below south west)},
            anchor=north west,
            yshift=0cm,
            xtick={0,0.05,0.1,0.15,0.2,0.25,0.3,0.35,0.4,0.45,0.5},
            xticklabel style={/pgf/number format/fixed},
            ytick={0,0.4,0.5,0.7,1},
            tick label style={font=\footnotesize},
            legend pos=north west,
            legend style={font=\scriptsize, fill=white, fill opacity=0.8, line width=0.3pt,inner sep=1pt,  
    outer sep=0pt},
            enlarge x limits=false,
            enlarge y limits=false,
        ]
            \addplot[black, dotted] coordinates {(0, 0.577) (0.6, 0.577)};
            \addplot[darklavender, solid] table[x=t, y=s1_i, col sep=comma] {data/Falla_AG_bivec_convert.csv};
            \addplot[darktangerine, solid] table[x=t, y=s2_i, col sep=comma] {data/Falla_AG_bivec_convert.csv};
            \addplot[bostonuniversityred, solid] table[x=t, y=s3_i, col sep=comma] {data/Falla_AG_bivec_convert.csv};

            \legend{Ref.,$\sigma_{12}$, $\sigma_{23}$, $\sigma_{31}$}
        \end{axis}  
    \end{tikzpicture}
    \caption{Three-phase voltage and current waveforms during line to ground (AG) fault event and voltage and current bivectors, simulated in the grid with converters.Pex}
    \label{fig:FAG}
\end{figure}

For line-to-ground (L-G) faults, it suffices to analyze the coefficients of the bivector within the plane of the curve. Fig.\ref{fig:FAG} illustrates these parameters for voltage and current in the Pex converter during an AG fault. Once the fault occurs, indicated at $t = 0.13$ s, the coefficients vary with respect to the reference point, and the voltage changes are depicted in Fig.\ref{fig:fault_classification}. Similarly, the BG and CG scenarios produce results that can be obtained with the interchanged coefficients, as shown in Fig.\ref{fig:fault_classification}.

\subsubsection{Line-to-line faults}

For line-to-line faults, the bivector coefficients remain constant within the Kirchhoff plane; consequently, their temporal representation is not pertinent. The classification is carried out solely based on the ellipse parameters.

\begin{figure}[tb]
    \centering
    \includegraphics[width=0.48\columnwidth]{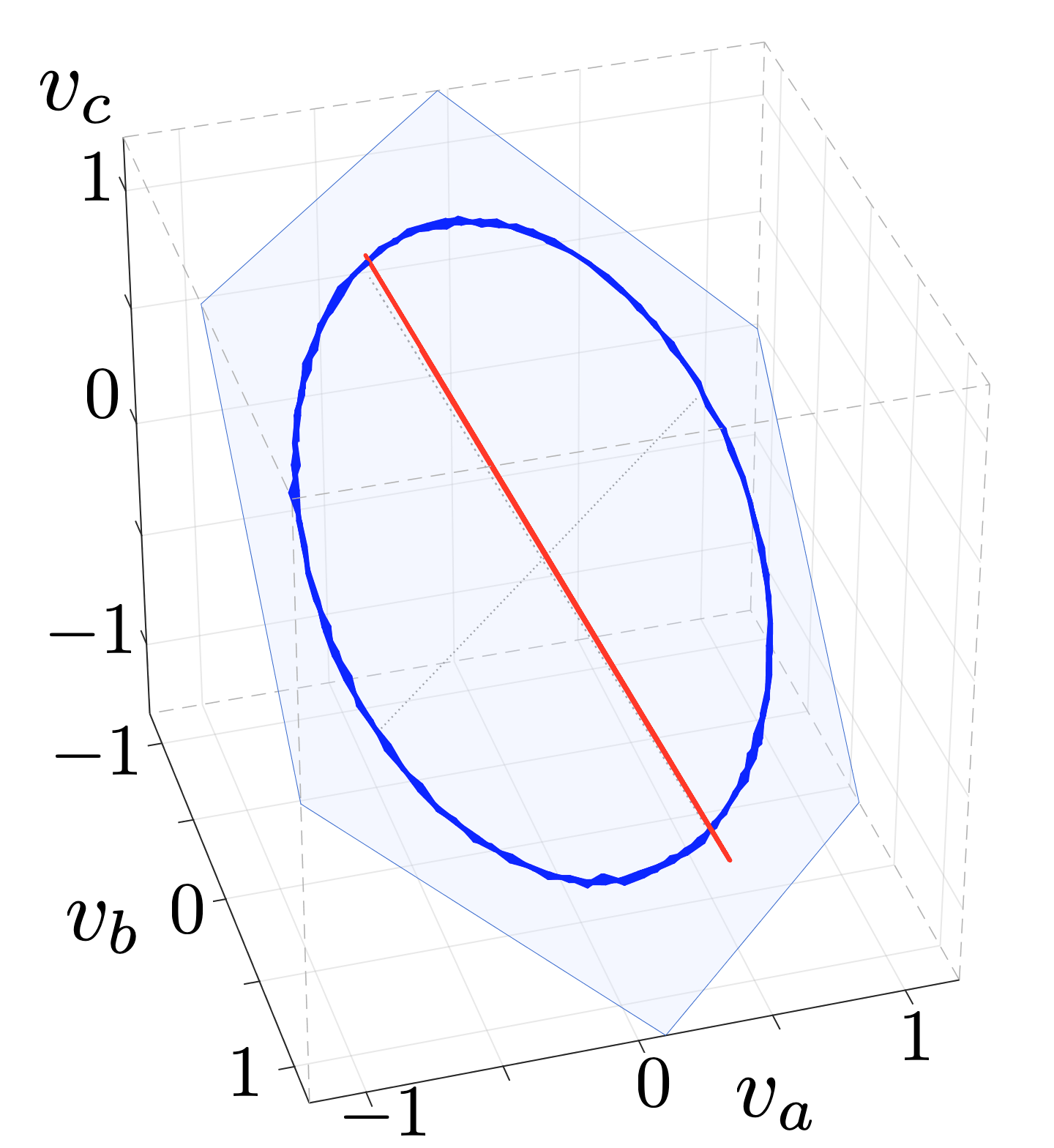}
    \includegraphics[width=0.48\columnwidth]{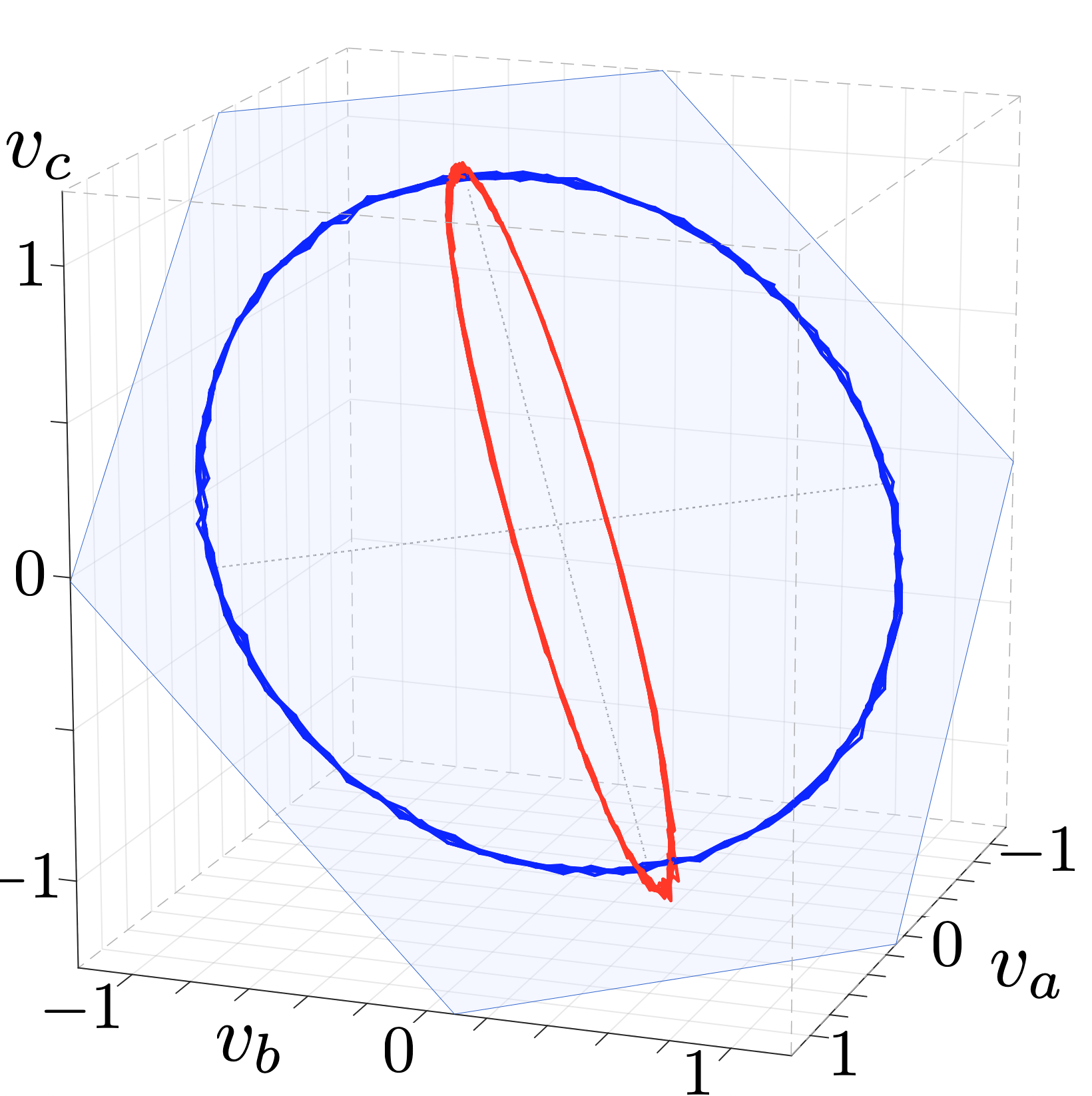}
    \caption{Different views of the voltage curve at Pex node 4 for AB fault using  (left) a pure resistor  and (right) a resistor plus inductor.}
    \label{fig:R&RL}
\end{figure}

Table \ref{tab:semiaxes_variations} shows that in the case of L-L faults, it is possible that the minor semi-axis equals 0. Therefore, the curve becomes a straight line, i.e. a degenerate ellipse, see Fig.\ref{fig:R&RL} (left). This situation occurs when the fault location is very close to the converter or the common coupling point, and the fault is resistive. For this type of fault, the two affected line voltages are equal, resulting in a straight line in the 3D geometric space on the Kirchhoff plane. When there is considerable reactance between these two points, a phase shift between the voltages arises, altering the curve's geometry to form an ellipse, as shown in Fig.\ref{fig:R&RL} (right). 

\begin{figure}[]
    \centering
    \begin{tikzpicture}
        \begin{axis}[
            width=0.99\columnwidth,
            height=3cm,
            xmin=0, xmax=0.28,
            ymin=-0.1, ymax=1.75,
            ylabel={Voltage\\ Semiaxes},
            ylabel style={font=\footnotesize, align=center},
            grid=major,
            grid style={line width=0.5pt, gray!50},
            line width=1pt,
            xticklabel=\empty,
            xtick={0,0.05,0.1,0.15,0.2,0.25,0.3,0.35,0.4,0.45},
            ytick={0,1.75,3},
            tick label style={font=\footnotesize},
            legend pos=north west,
            legend style={font=\footnotesize, fill=white, fill opacity=0.8, line width=0.3pt},
            legend columns=2,
            legend image post style={xscale=0.5},
            name=voltage_plot1,
            enlarge x limits=false,
            enlarge y limits=false,
        ]
            \addplot[blue, solid] table[x=t, y=R_v, col sep=comma] {data/Falla_A-B_Pex_ellip.csv};  
            \addplot[blue, dotted] table[x=t, y=r_v, col sep=comma] {data/Falla_A-B_Pex_ellip.csv};  
            
            \addplot[red, solid] table[x=t, y=R_v, col sep=comma] {data/Falla_B-C_Pex_ellip.csv};  
            \addplot[red, dotted] table[x=t, y=r_v, col sep=comma] {data/Falla_B-C_Pex_ellip.csv};  
            
            \addplot[black, solid] table[x=t, y=R_v, col sep=comma] {data/Falla_A-C_Pex_ellip.csv};  
            \addplot[black, dotted] table[x=t, y=r_v, col sep=comma] {data/Falla_A-C_Pex_ellip.csv}; 
            \legend{$ a_{\scalebox{0.7}{AB}}$,$ b_{\scalebox{0.7}{AB}}$,$a_{\scalebox{0.7}{BC}}$,$b_{\scalebox{0.7}{BC}}$,$a_{\scalebox{0.7}{CA}}$,$b_{\scalebox{0.7}{CA}}$}
        \end{axis}

        \begin{axis}[
            width=0.99\columnwidth,
            height=3cm,
            xmin=0, xmax=0.28,
            ymin=-0.02, ymax=3.6,
            xlabel style={font=\footnotesize},
            ylabel={Current\\ Semiaxes}, 
            ylabel style={font=\footnotesize, align=center},
            grid=major,
            grid style={line width=0.5pt, gray!50},
            line width=1pt,
            at={(voltage_plot1.below south west)},
            anchor=north west,
            yshift=0cm,
            xticklabel=\empty,
            xtick={0,0.05,0.1,0.15,0.2,0.25,0.3,0.35,0.4,0.45},
            ytick={0,1.75,3.5},
            tick label style={font=\footnotesize},
            legend pos=north west,
            legend style={font=\footnotesize, fill=white, fill opacity=0.8, line width=0.3pt},
            legend columns=2,
            legend image post style={xscale=0.5},
            name=voltage_plot2, 
            enlarge x limits=false,
            enlarge y limits=false,
        ]
            \addplot[blue, solid] table[x=t, y=R_i, col sep=comma] {data/Falla_A-B_Pex_ellip.csv}; 
            \addplot[blue, dotted] table[x=t, y=r_i, col sep=comma] {data/Falla_A-B_Pex_ellip.csv};
            
            \addplot[red, solid] table[x=t, y=R_i, col sep=comma] {data/Falla_B-C_Pex_ellip.csv}; 
            \addplot[red, dotted] table[x=t, y=r_i, col sep=comma] {data/Falla_B-C_Pex_ellip.csv};  
            
            \addplot[black, solid] table[x=t, y=R_i, col sep=comma] {data/Falla_A-C_Pex_ellip.csv};
            \addplot[black, dotted] table[x=t, y=r_i, col sep=comma] {data/Falla_A-C_Pex_ellip.csv}; 
            \legend{$ a_{\scalebox{0.7}{AB}}$,$ b_{\scalebox{0.7}{AB}}$,$a_{\scalebox{0.7}{BC}}$,$b_{\scalebox{0.7}{BC}}$,$a_{\scalebox{0.7}{CA}}$,$b_{\scalebox{0.7}{CA}}$}
        \end{axis}

        \begin{axis}[
            width=0.99\columnwidth,
            height=3cm,
            xmin=0, xmax=0.28,
            ymin=-0.1, ymax=3.5,
            ylabel={Voltage\\Angle (rad)},
            ylabel style={font=\footnotesize, align=center},
            grid=major,
            grid style={line width=0.5pt, gray!50},
            line width=1pt,
            at={(voltage_plot2.below south west)},
            anchor=north west,
            yshift=0cm,
            xticklabel=\empty,
            xtick={0,0.05,0.1,0.15,0.2,0.25,0.3,0.35,0.4,0.45},
            ytick={0,1.75,3},
            tick label style={font=\footnotesize},
            legend pos=north west,
            legend style={font=\footnotesize, fill=white, fill opacity=0.8, line width=0.3pt},
            name=voltage_plot3,
            enlarge x limits=false,
            enlarge y limits=false,
        ]
                    
            \fill[red!30, opacity=0.5] (axis cs:0.12,0.2618) rectangle (axis cs:0.3,1.3090);
            \fill[green!30, opacity=0.5] (axis cs:0.12,1.3090) rectangle (axis cs:0.3,2.3562);
            \fill[blue!30, opacity=0.5] (axis cs:0.12,2.3562) rectangle (axis cs:0.3,3.4034);
            
            \addplot[blue, solid] table[x=t, y=a_v, col sep=comma] {data/Falla_A-B_Pex_ellip.csv};  \addplot[red, solid] table[x=t, y=a_v, col sep=comma] {data/Falla_B-C_Pex_ellip.csv};  \addplot[black, solid] table[x=t, y=a_v, col sep=comma] {data/Falla_A-C_Pex_ellip.csv};      
            
            \legend{$\theta_{\scalebox{0.7}{AB}}$, $\theta_{\scalebox{0.7}{BC}}$,$\theta_{\scalebox{0.7}{CA}}$}
        \end{axis}

        \begin{axis}[
            width=0.99\columnwidth,
            height=3cm,
            xmin=0, xmax=0.28,
            ymin=-0.02, ymax=3.5,
            xlabel={Time (s)},
            xlabel style={font=\footnotesize},
            ylabel={Current\\Angle (rad)}, 
            ylabel style={font=\footnotesize, align=center},
            grid=major,
            grid style={line width=0.5pt, gray!50},
            line width=1pt,
            at={(voltage_plot3.below south west)},
            anchor=north west,
            yshift=0cm,
            xtick={0,0.05,0.1,0.15,0.2,0.25,0.3,0.35,0.4,0.45},
            xticklabel style={/pgf/number format/fixed},
            ytick={0,1.75,3.5},
            tick label style={font=\footnotesize},
            legend pos=north west,
            legend style={font=\footnotesize, fill=white, fill opacity=0.8, line width=0.3pt},
            name=voltage_plot4, 
            enlarge x limits=false,
            enlarge y limits=false,
        ]
            \addplot[blue, solid] table[x=t, y=a_i, col sep=comma] {data/Falla_A-B_Pex_ellip.csv};  \addplot[red, solid] table[x=t, y=a_i, col sep=comma] {data/Falla_B-C_Pex_ellip.csv};  
            \addplot[black, solid] table[x=t, y=a_i, col sep=comma] {data/Falla_A-C_Pex_ellip.csv};          
            \legend{$\theta_{\scalebox{0.7}{AB}}$, $\theta_{\scalebox{0.7}{BC}}$,$\theta_{\scalebox{0.7}{CA}}$}
        \end{axis}

    \end{tikzpicture}
    \caption{Ellipse parameters over time for voltage and current curves during LLF event in the Pex (GFL) converter with $R$ fault.}
    \label{fig:FLL_Pex_R}
\end{figure}
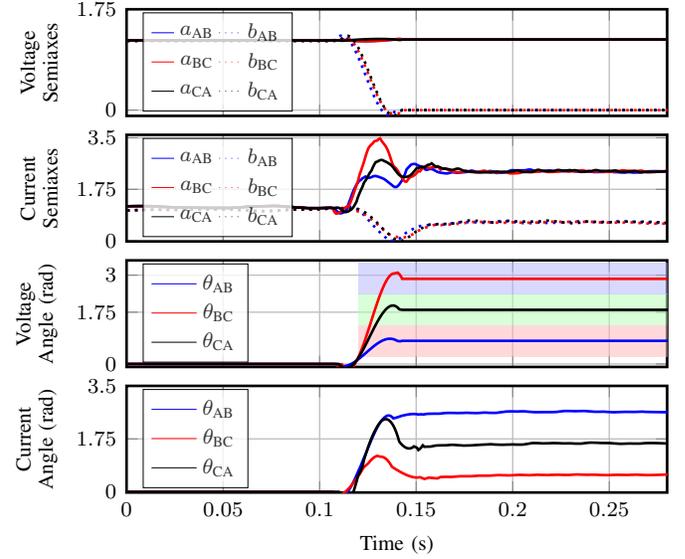

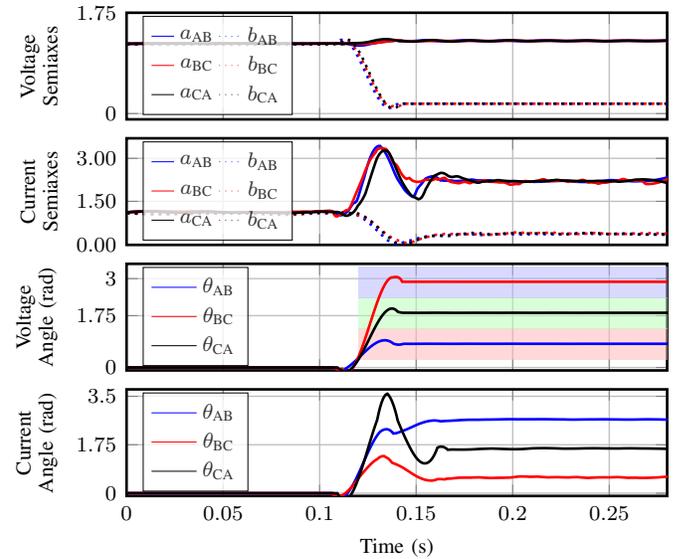
\begin{figure}[]
    \centering
    \begin{tikzpicture}
        \begin{axis}[
            width=0.99\columnwidth,
            height=3cm,
            xmin=0, xmax=0.28,
            ymin=-0.1, ymax=1.75,
            ylabel={Voltage\\ Semiaxes},
            ylabel style={font=\footnotesize, align=center},
            grid=major,
            grid style={line width=0.5pt, gray!50},
            line width=1pt,
            xticklabel=\empty,
            xtick={0,0.05,0.1,0.15,0.2,0.25,0.3,0.35,0.4,0.45},
            ytick={0,1.75,3},
            tick label style={font=\footnotesize},
            legend pos=north west,
            legend style={font=\footnotesize, fill=white, fill opacity=0.8, line width=0.3pt},
            legend columns=2,
            legend image post style={xscale=0.5},
            name=voltage_plot1,
            enlarge x limits=false,
            enlarge y limits=false,
        ]
            \addplot[blue, solid] table[x=t, y=R_v, col sep=comma] {data/FallaRL_A-B_Pex_ellip.csv};  
            \addplot[blue, dotted] table[x=t, y=r_v, col sep=comma] {data/FallaRL_A-B_Pex_ellip.csv}; 
            
            \addplot[red, solid] table[x=t, y=R_v, col sep=comma] {data/FallaRL_B-C_Pex_ellip.csv}; 
            \addplot[red, dotted] table[x=t, y=r_v, col sep=comma] {data/FallaRL_B-C_Pex_ellip.csv};  
            
            \addplot[black, solid] table[x=t, y=R_v, col sep=comma] {data/FallaRL_A-C_Pex_ellip.csv};  
            \addplot[black, dotted] table[x=t, y=r_v, col sep=comma] {data/FallaRL_A-C_Pex_ellip.csv}; 
            \legend{$ a_{\scalebox{0.7}{AB}}$,$ b_{\scalebox{0.7}{AB}}$,$a_{\scalebox{0.7}{BC}}$,$b_{\scalebox{0.7}{BC}}$,$a_{\scalebox{0.7}{CA}}$,$b_{\scalebox{0.7}{CA}}$}
        \end{axis}

        \begin{axis}[
            width=0.99\columnwidth,
            height=3cm,
            xmin=0, xmax=0.28,
            ymin=-0.02, ymax=3.7,
            xlabel style={font=\footnotesize},
            ylabel={Current\\ Semiaxes}, 
            ylabel style={font=\footnotesize, align=center},
            grid=major,
            grid style={line width=0.5pt, gray!50},
            line width=1pt,
            at={(voltage_plot1.below south west)},
            anchor=north west,
            yshift=0cm,
            xticklabel=\empty,
            xtick={0,0.05,0.1,0.15,0.2,0.25,0.3,0.35,0.4,0.45},
            ytick={0,1.5,3},
            yticklabel={\num[round-mode=places, round-precision=2]{\tick}},
            tick label style={font=\footnotesize},
            legend pos=north west,
            legend style={font=\footnotesize, fill=white, fill opacity=0.8, line width=0.3pt},
            legend columns=2,
            legend image post style={xscale=0.5},
            name=voltage_plot2, 
            enlarge x limits=false,
            enlarge y limits=false,
        ]
            \addplot[blue, solid] table[x=t, y=R_i, col sep=comma] {data/FallaRL_A-B_Pex_ellip.csv}; 
            \addplot[blue, dotted] table[x=t, y=r_i, col sep=comma] {data/FallaRL_A-B_Pex_ellip.csv};  
            
            \addplot[red, solid] table[x=t, y=R_i, col sep=comma] {data/FallaRL_B-C_Pex_ellip.csv}; 
            \addplot[red, dotted] table[x=t, y=r_i, col sep=comma] {data/FallaRL_B-C_Pex_ellip.csv};
            
            \addplot[black, solid] table[x=t, y=R_i, col sep=comma] {data/FallaRL_A-C_Pex_ellip.csv}; 
            \addplot[black, dotted] table[x=t, y=r_i, col sep=comma] {data/FallaRL_A-C_Pex_ellip.csv}; 
            \legend{$ a_{\scalebox{0.7}{AB}}$,$ b_{\scalebox{0.7}{AB}}$,$a_{\scalebox{0.7}{BC}}$,$b_{\scalebox{0.7}{BC}}$,$a_{\scalebox{0.7}{CA}}$,$b_{\scalebox{0.7}{CA}}$}
        \end{axis}

        \begin{axis}[
            width=0.99\columnwidth,
            height=3cm,
            xmin=0, xmax=0.28,
            ymin=-0.1, ymax=3.5,
            ylabel={Voltage\\Angle (rad)},
            ylabel style={font=\footnotesize, align=center},
            grid=major,
            grid style={line width=0.5pt, gray!50},
            line width=1pt,
            at={(voltage_plot2.below south west)},
            anchor=north west,
            yshift=0cm,
            xticklabel=\empty,
            xtick={0,0.05,0.1,0.15,0.2,0.25,0.3,0.35,0.4,0.45},
            ytick={0,1.75,3},
            tick label style={font=\footnotesize},
            legend pos=north west,
            legend style={font=\footnotesize, fill=white, fill opacity=0.8, line width=0.3pt},
            name=voltage_plot3,
            enlarge x limits=false,
            enlarge y limits=false,
        ]
            \fill[red!30, opacity=0.5] (axis cs:0.12,0.2618) rectangle (axis cs:0.3,1.3090);
            \fill[green!30, opacity=0.5] (axis cs:0.12,1.3090) rectangle (axis cs:0.3,2.3562);
            \fill[blue!30, opacity=0.5] (axis cs:0.12,2.3562) rectangle (axis cs:0.3,3.4034);
            
            \addplot[blue, solid] table[x=t, y=a_v, col sep=comma] {data/FallaRL_A-B_Pex_ellip.csv};  \addplot[red, solid] table[x=t, y=a_v, col sep=comma] {data/FallaRL_B-C_Pex_ellip.csv};  \addplot[black, solid] table[x=t, y=a_v, col sep=comma] {data/FallaRL_A-C_Pex_ellip.csv};      
            \legend{$\theta_{\scalebox{0.7}{AB}}$, $\theta_{\scalebox{0.7}{BC}}$,$\theta_{\scalebox{0.7}{CA}}$}
        \end{axis}

        \begin{axis}[
            width=0.99\columnwidth,
            height=3cm,
            xmin=0, xmax=0.28,
            ymin=-0.1, ymax=3.75,
            xlabel={Time (s)},
            xlabel style={font=\footnotesize},
            ylabel={Current\\Angle (rad)}, 
            ylabel style={font=\footnotesize, align=center},
            grid=major,
            grid style={line width=0.5pt, gray!50},
            line width=1pt,
            at={(voltage_plot3.below south west)},
            anchor=north west,
            yshift=0cm,
            xtick={0,0.05,0.1,0.15,0.2,0.25,0.3,0.35,0.4,0.45},
            xticklabel style={/pgf/number format/fixed},
            ytick={0,1.75,3.5},
            tick label style={font=\footnotesize},
            legend pos=north west,
            legend style={font=\footnotesize, fill=white, fill opacity=0.8, line width=0.3pt},
            name=voltage_plot4, 
            enlarge x limits=false,
            enlarge y limits=false,
        ]
            \addplot[blue, solid] table[x=t, y=a_i, col sep=comma] {data/FallaRL_A-B_Pex_ellip.csv};  \addplot[red, solid] table[x=t, y=a_i, col sep=comma] {data/FallaRL_B-C_Pex_ellip.csv};  
            \addplot[black, solid] table[x=t, y=a_i, col sep=comma] {data/FallaRL_A-C_Pex_ellip.csv};          
            \legend{$\theta_{\scalebox{0.7}{AB}}$, $\theta_{\scalebox{0.7}{BC}}$,$\theta_{\scalebox{0.7}{CA}}$}
        \end{axis}

    \end{tikzpicture}
    \caption{Ellipse parameters over time for voltage and current curves during LLF event in the Pex (GFL) converter with $RL$ fault.}
    \label{fig:FLL_Pex_RL}
\end{figure}

The method for identifying and addressing failures with degenerate ellipses involves a simple variation of the existing approach. When the curve is a straight line, the magnitude of the bivector $\bm{B}$ approaches zero, as the initial and final vectors of the analysis window are nearly parallel. The curve remains in the Kirchhoff plane, and after applying the transformation that rotates it into the plane $\bm{\sigma_{12}}$, the fitting is adjusted to a line using least squares, resulting in the angle of inclination and the length of the line of the faulted phases. Figs. \ref{fig:FLL_Pex_R} and \ref{fig:FLL_Pex_RL} show the results of the method for the cases of resistive and resistive plus reactive impedances between the faulted point and the Pex converter, respectively.

For the voltage analysis, it is observed that the minor semi-axis in the purely resistive fault impedance configuration ($R$) decreases to zero for each fault type in Fig \ref{fig:FLL_Pex_R}. In the resistive-inductive fault ($RL$) the minor semi-axis is greater than zero. However, the ellipse inclination angles reach identical values in both fault configurations. Hence, this enables accurate fault classification according to the sectoral diagram presented in Fig. \ref{fig:fault_classification_circle}. The current analysis demonstrates that the semi-axes variations are more pronounced for the $RL$ fault impedance configuration. $R$ and $RL$ configurations yield consistent ellipse inclination angles, as demonstrated in Figs. \ref{fig:FLL_Pex_R} and \ref{fig:FLL_Pex_RL}.

\begin{figure}[]
    \centering
    \begin{tikzpicture}
        \begin{axis}[
            width=0.99\columnwidth,
            height=3cm,
            xmin=0, xmax=0.28,
            ymin=1.14, ymax=1.34,
            ylabel={Voltage\\ Semiaxes},
            ylabel style={font=\footnotesize, align=center},
            grid=major,
            grid style={line width=0.5pt, gray!50},
            line width=1pt,
            xticklabel=\empty,
            xtick={0,0.05,0.1,0.15,0.2,0.25,0.3,0.35,0.4,0.45},
            ytick={0,1,1.15,1.2,1.25,1.3},
            yticklabel={$\phantom{3}$\num[round-mode=places, round-precision=2]{\tick}},
            tick label style={font=\footnotesize},
            legend pos=north west,
            legend style={font=\footnotesize, fill=white, fill opacity=0.8, line width=0.3pt},
            legend columns=2,
            legend image post style={xscale=0.5},
            name=voltage_plot1,
            enlarge x limits=false,
            enlarge y limits=false,
        ]
            \addplot[blue, solid] table[x=t, y=R_v, col sep=comma] {data/Falla_A-B_DG1_ellip.csv}; 
            \addplot[blue, dotted] table[x=t, y=r_v, col sep=comma] {data/Falla_A-B_DG1_ellip.csv}; 
            
            \addplot[red, solid] table[x=t, y=R_v, col sep=comma] {data/Falla_B-C_DG1_ellip.csv};  
            \addplot[red, dotted] table[x=t, y=r_v, col sep=comma] {data/Falla_B-C_DG1_ellip.csv};
            
            \addplot[black, solid] table[x=t, y=R_v, col sep=comma] {data/Falla_A-C_DG1_ellip.csv};   
            \addplot[black, dotted] table[x=t, y=r_v, col sep=comma] {data/Falla_A-C_DG1_ellip.csv}; 
            \legend{$a_{\scalebox{0.7}{AB}}$,$b_{\scalebox{0.7}{AB}}$,$a_{\scalebox{0.7}{BC}}$,$b_{\scalebox{0.7}{BC}}$,$a_{\scalebox{0.7}{CA}}$,$b_{\scalebox{0.7}{CA}}$}
        \end{axis}

        \begin{axis}[
            width=0.99\columnwidth,
            height=3cm,
            xmin=0, xmax=0.28,
            ymin=-0.02, ymax=30,
            xlabel style={font=\footnotesize},
            ylabel={Current\\ Semiaxes}, 
            ylabel style={font=\footnotesize, align=center},
            grid=major,
            grid style={line width=0.5pt, gray!50},
            line width=1pt,
            at={(voltage_plot1.below south west)},
            anchor=north west,
            yshift=0cm,
            xticklabel=\empty,
            xtick={0,0.05,0.1,0.15,0.2,0.25,0.3,0.35,0.4,0.45},
            ytick={0,10.0,20.0,30.0},
            yticklabel={\num[round-mode=places, round-precision=2]{\tick}},
            tick label style={font=\footnotesize},
            legend pos=north west,
            legend style={font=\footnotesize, fill=white, fill opacity=0.8, line width=0.3pt},
            legend columns=2,
            legend image post style={xscale=0.5},
            name=voltage_plot2, 
            enlarge x limits=false,
            enlarge y limits=false,
        ]
            \addplot[blue, solid] table[x=t, y=R_i, col sep=comma] {data/Falla_A-B_DG1_ellip.csv};  
            \addplot[blue, dotted] table[x=t, y=r_i, col sep=comma] {data/Falla_A-B_DG1_ellip.csv};
            
            \addplot[red, solid] table[x=t, y=R_i, col sep=comma] {data/Falla_B-C_DG1_ellip.csv};
            \addplot[red, dotted] table[x=t, y=r_i, col sep=comma] {data/Falla_B-C_DG1_ellip.csv};  
            
            \addplot[black, solid] table[x=t, y=R_i, col sep=comma] {data/Falla_A-C_DG1_ellip.csv};
            \addplot[black, dotted] table[x=t, y=r_i, col sep=comma] {data/Falla_A-C_DG1_ellip.csv}; 
            \legend{$ a_{\scalebox{0.7}{AB}}$,$ b_{\scalebox{0.7}{AB}}$,$a_{\scalebox{0.7}{BC}}$,$b_{\scalebox{0.7}{BC}}$,$a_{\scalebox{0.7}{CA}}$,$b_{\scalebox{0.7}{CA}}$}
        \end{axis}

        \begin{axis}[
            width=0.99\columnwidth,
            height=3cm,
            xmin=0, xmax=0.28,
            ymin=-0.1, ymax=3.5,
            ylabel={Voltage\\Angle (rad)},
            ylabel style={font=\footnotesize, align=center},
            grid=major,
            grid style={line width=0.5pt, gray!50},
            line width=1pt,
            at={(voltage_plot2.below south west)},
            anchor=north west,
            yshift=0cm,
            xticklabel=\empty,
            xtick={0,0.05,0.1,0.15,0.2,0.25,0.3,0.35,0.4,0.45},
            ytick={0,1.50,3.0},
            yticklabel={$\phantom{1}$\num[round-mode=places, round-precision=2]{\tick}},
            tick label style={font=\footnotesize},
            legend pos=north west,
            legend style={font=\footnotesize, fill=white, fill opacity=0.8, line width=0.3pt},
            name=voltage_plot3,
            enlarge x limits=false,
            enlarge y limits=false,
        ]
            \fill[red!30, opacity=0.5] (axis cs:0.12,0.2618) rectangle (axis cs:0.3,1.3090);
            \fill[green!30, opacity=0.5] (axis cs:0.12,1.3090) rectangle (axis cs:0.3,2.3562);
            \fill[blue!30, opacity=0.5] (axis cs:0.12,2.3562) rectangle (axis cs:0.3,3.4034);
            
            \addplot[blue, solid] table[x=t, y=a_v, col sep=comma] {data/Falla_A-B_DG1_ellip.csv};  \addplot[red, solid] table[x=t, y=a_v, col sep=comma] {data/Falla_B-C_DG1_ellip.csv};  \addplot[black, solid] table[x=t, y=a_v, col sep=comma] {data/Falla_A-C_DG1_ellip.csv};      
            \legend{$\theta_{\scalebox{0.7}{AB}}$, $\theta_{\scalebox{0.7}{BC}}$,$\theta_{\scalebox{0.7}{CA}}$}
        \end{axis}

        \begin{axis}[
            width=0.99\columnwidth,
            height=3cm,
            xmin=0, xmax=0.28,
            ymin=-0.02, ymax=3.5,
            xlabel={Time (s)},
            xlabel style={font=\footnotesize},
            ylabel={Current\\Angle (rad)}, 
            ylabel style={font=\footnotesize, align=center},
            grid=major,
            grid style={line width=0.5pt, gray!50},
            line width=1pt,
            at={(voltage_plot3.below south west)},
            anchor=north west,
            yshift=0cm,
            xtick={0,0.05,0.1,0.15,0.2,0.25,0.3,0.35,0.4,0.45},
            xticklabel style={/pgf/number format/fixed},
            ytick={0,1.75,3.5},
            yticklabel={$\phantom{1}$\num[round-mode=places, round-precision=2]{\tick}},
            tick label style={font=\footnotesize},
            legend pos=north west,
            legend style={font=\footnotesize, fill=white, fill opacity=0.8, line width=0.3pt},
            name=voltage_plot4, 
            enlarge x limits=false,
            enlarge y limits=false,
        ]
            \addplot[blue, solid] table[x=t, y=a_i, col sep=comma] {data/Falla_A-B_DG1_ellip.csv};  \addplot[red, solid] table[x=t, y=a_i, col sep=comma] {data/Falla_B-C_DG1_ellip.csv};  
            \addplot[black, solid] table[x=t, y=a_i, col sep=comma] {data/Falla_A-C_DG1_ellip.csv};          
            \legend{$\theta_{\scalebox{0.7}{AB}}$, $\theta_{\scalebox{0.7}{BC}}$,$\theta_{\scalebox{0.7}{CA}}$}
        \end{axis}
 
    \end{tikzpicture}
    \caption{Ellipse parameters over time for voltage and current curve ellipse fitting during LLF event, DG1 (GFM).}
    \label{fig:FLL_DG1}
\end{figure}
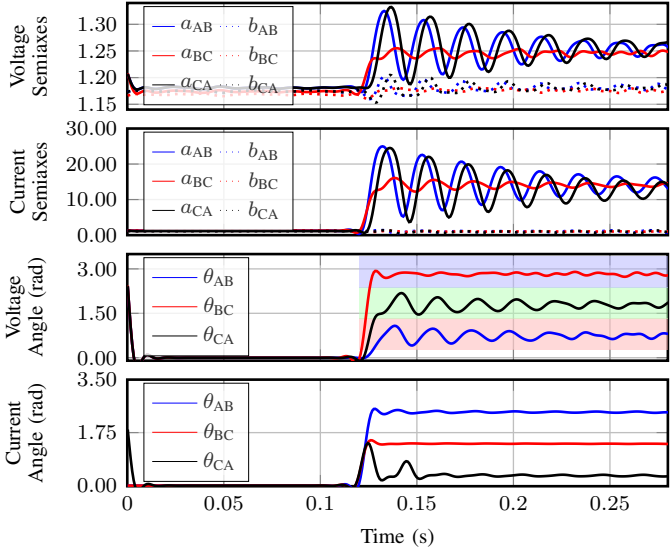

Fig. \ref{fig:FLL_DG1} shows major and minor axes with oscillatory behavior in the case of DG1 with GFM control. The controller strategy is clearly different that the GFL, resulting in the voltage and current curves with elliptical geometries displaced from the origin. Consequently, the fitting method yields parameters with some oscillation. However, the oscillations do not hinder the identification of the type of fault. Despite the oscillation, the three bands of inclination angles that characterize the faults are clearly identified. The same phenomenon occurs with the current; however, the values do not coincide with the Fig. \ref{fig:FLL_Pex_R} and Fig. \ref{fig:FLL_Pex_RL} of Pex due to the use of different control strategies.

\subsection{Experimental validation}

\begin{figure}[]
	   \centering 
        \includegraphics[width=\columnwidth]{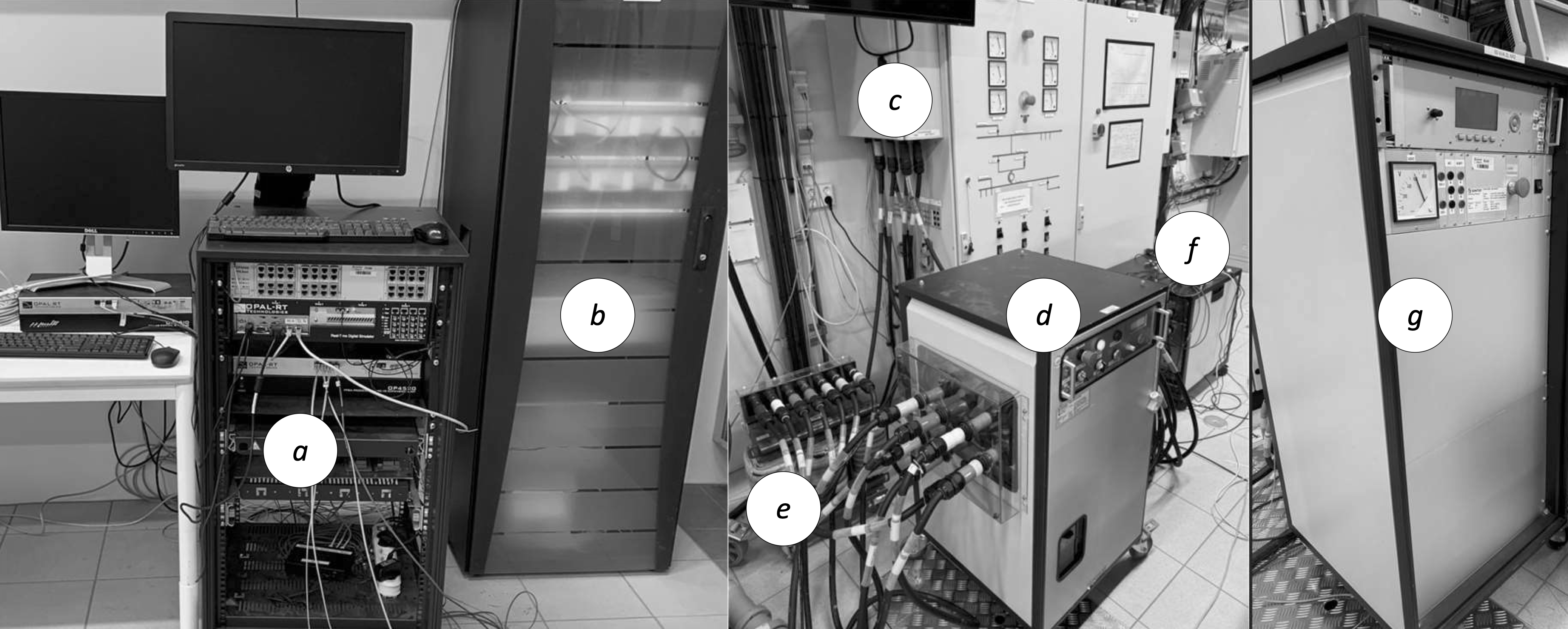}
	    \caption{Experimental setup. a) OPAL-RT, b) optic fiber communications, c) main grid, d) Fault emulator device, e) Reactance, f) measurement box and g) 60 kVA voltage source converter.}
	    \label{fig:nsgl}
	\end{figure}

 \begin{figure}[]
        \centering
        \includegraphics[width=1\linewidth]{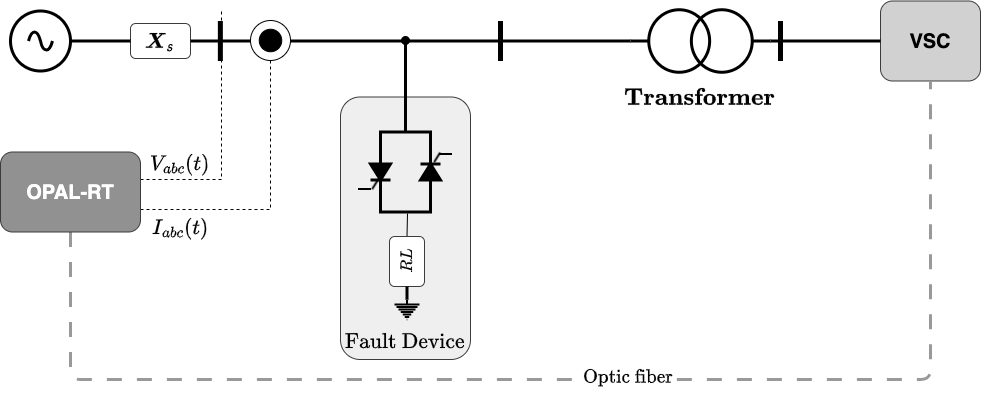}
        \caption{Scheme of the circuit used in the experimental validation.}
        \label{fig:schemaLab}
    \end{figure}

The laboratory setup shown in Fig. \ref{fig:nsgl} has been used to collect physical signals in a representative scenario with a 60 kVA IBR and a fault emulator. The configuration of the devices is presented in Fig. \ref{fig:schemaLab}. The data from the laboratory have asymmetric LLFs and the symmetric three-phase fault. The signals were captured every 100 $\mu$s. This was achieved using a converter control with an OPAL-RT real-time simulator.

\begin{figure}[]
    \centering
    \begin{tikzpicture}
        \begin{axis}[
            width=0.99\columnwidth,
            height=3cm,
            xmin=0, xmax=0.28,
            ymin=-0.1, ymax=1.75,
            ylabel={Voltage\\ Semiaxes},
            ylabel style={font=\footnotesize, align=center},
            grid=major,
            grid style={line width=0.5pt, gray!50},
            line width=1pt,
            xticklabel=\empty,
            xtick={0,0.05,0.1,0.15,0.2,0.25,0.3,0.35,0.4,0.45},
            ytick={0,1.75,3},
            tick label style={font=\footnotesize},
            legend pos=north west,
            legend style={font=\footnotesize, fill=white, fill opacity=0.8, line width=0.3pt},
            legend columns=2,
            legend image post style={xscale=0.5},
            name=voltage_plot1,
            enlarge x limits=false,
            enlarge y limits=false,
        ]
            \addplot[blue, solid] table[x=t, y=R_v, col sep=comma] {data/Falla_A-B_ellip.csv};  
            \addplot[blue, dotted] table[x=t, y=r_v, col sep=comma] {data/Falla_A-B_ellip.csv}; 
            
            \addplot[red, solid] table[x=t, y=R_v, col sep=comma] {data/Falla_B-C_ellip.csv};  
            \addplot[red, dotted] table[x=t, y=r_v, col sep=comma] {data/Falla_B-C_ellip.csv}; 
            
            \addplot[black, solid] table[x=t, y=R_v, col sep=comma] {data/Falla_A-C_ellip.csv};  
            \addplot[black, dotted] table[x=t, y=r_v, col sep=comma] {data/Falla_A-C_ellip.csv}; 
            \legend{$ a_{\scalebox{0.7}{AB}}$,$ b_{\scalebox{0.7}{AB}}$,$a_{\scalebox{0.7}{BC}}$,$b_{\scalebox{0.7}{BC}}$,$a_{\scalebox{0.7}{CA}}$,$b_{\scalebox{0.7}{CA}}$}
        \end{axis}

        \begin{axis}[
            width=0.99\columnwidth,
            height=3cm,
            xmin=0, xmax=0.28,
            ymin=-0.02, ymax=9,
            xlabel style={font=\footnotesize},
            ylabel={Current\\ Semiaxes}, 
            ylabel style={font=\footnotesize, align=center},
            grid=major,
            grid style={line width=0.5pt, gray!50},
            line width=1pt,
            at={(voltage_plot1.below south west)},
            anchor=north west,
            yshift=0cm,
            xticklabel=\empty,
            xtick={0,0.05,0.1,0.15,0.2,0.25,0.3,0.35,0.4,0.45},
            ytick={0,4.5,9},
            yticklabel={\num[round-mode=places, round-precision=2]{\tick}},
            tick label style={font=\footnotesize},
            legend pos=north west,
            legend style={font=\footnotesize, fill=white, fill opacity=0.8, line width=0.3pt},
            legend columns=2,
            legend image post style={xscale=0.5},
            name=voltage_plot2, 
            enlarge x limits=false,
            enlarge y limits=false,
        ]
            \addplot[blue, solid] table[x=t, y=R_i, col sep=comma] {data/Falla_A-B_ellip.csv};  
            \addplot[blue, dotted] table[x=t, y=r_i, col sep=comma] {data/Falla_A-B_ellip.csv};  
            
            \addplot[red, solid] table[x=t, y=R_i, col sep=comma] {data/Falla_B-C_ellip.csv};  
            \addplot[red, dotted] table[x=t, y=r_i, col sep=comma] {data/Falla_B-C_ellip.csv}; 
            
            \addplot[black, solid] table[x=t, y=R_i, col sep=comma] {data/Falla_A-C_ellip.csv};  
            \addplot[black, dotted] table[x=t, y=r_i, col sep=comma] {data/Falla_A-C_ellip.csv}; 
            \legend{$ a_{\scalebox{0.7}{AB}}$,$ b_{\scalebox{0.7}{AB}}$,$a_{\scalebox{0.7}{BC}}$,$b_{\scalebox{0.7}{BC}}$,$a_{\scalebox{0.7}{CA}}$,$b_{\scalebox{0.7}{CA}}$}
        \end{axis}

        \begin{axis}[
            width=0.99\columnwidth,
            height=3cm,
            xmin=0, xmax=0.28,
            ymin=-0.1, ymax=3.5,
            ylabel={Voltage\\Angle (rad)},
            ylabel style={font=\footnotesize, align=center},
            grid=major,
            grid style={line width=0.5pt, gray!50},
            line width=1pt,
            at={(voltage_plot2.below south west)},
            anchor=north west,
            yshift=0cm,
            xticklabel=\empty,
            xtick={0,0.05,0.1,0.15,0.2,0.25,0.3,0.35,0.4,0.45},
            ytick={0,1.75,3},
            tick label style={font=\footnotesize},
            legend pos=north west,
            legend style={font=\footnotesize, fill=white, fill opacity=0.8, line width=0.3pt},
            name=voltage_plot3,
            enlarge x limits=false,
            enlarge y limits=false,
        ]
            \fill[red!30, opacity=0.5] (axis cs:0.12,0.2618) rectangle (axis cs:0.3,1.3090);
            \fill[green!30, opacity=0.5] (axis cs:0.12,1.3090) rectangle (axis cs:0.3,2.3562);
            \fill[blue!30, opacity=0.5] (axis cs:0.12,2.3562) rectangle (axis cs:0.3,3.4034);
            
            \addplot[blue, solid] table[x=t, y=a_v, col sep=comma] {data/Falla_A-B_ellip.csv};  
            \addplot[red, solid] table[x=t, y=a_v, col sep=comma] {data/Falla_B-C_ellip.csv};  
            \addplot[black, solid] table[x=t, y=a_v, col sep=comma] {data/Falla_A-C_ellip.csv};      
            \legend{$\theta_{\scalebox{0.7}{AB}}$, $\theta_{\scalebox{0.7}{BC}}$,$\theta_{\scalebox{0.7}{CA}}$}
        \end{axis}

        \begin{axis}[
            width=0.99\columnwidth,
            height=3cm,
            xmin=0, xmax=0.28,
            ymin=-0.02, ymax=3.5,
            xlabel={Time (s)},
            xlabel style={font=\footnotesize},
            ylabel={Current\\Angle (rad)}, 
            ylabel style={font=\footnotesize, align=center},
            grid=major,
            grid style={line width=0.5pt, gray!50},
            line width=1pt,
            at={(voltage_plot3.below south west)},
            anchor=north west,
            yshift=0cm,
            xtick={0,0.05,0.1,0.15,0.2,0.25,0.3,0.35,0.4,0.45},
            xticklabel style={/pgf/number format/fixed},
            ytick={0,1.75,3.5},
            tick label style={font=\footnotesize},
            legend pos=north west,
            legend style={font=\footnotesize, fill=white, fill opacity=0.8, line width=0.3pt},
            name=voltage_plot4, 
            enlarge x limits=false,
            enlarge y limits=false,
        ]
            \addplot[blue, solid] table[x=t, y=a_i, col sep=comma] {data/Falla_A-B_ellip.csv};  \addplot[red, solid] table[x=t, y=a_i, col sep=comma] {data/Falla_B-C_ellip.csv};  
            \addplot[black, solid] table[x=t, y=a_i, col sep=comma] {data/Falla_A-C_ellip.csv};          
            \legend{$\theta_{\scalebox{0.7}{AB}}$, $\theta_{\scalebox{0.7}{BC}}$,$\theta_{\scalebox{0.7}{CA}}$}
        \end{axis}

    \end{tikzpicture}
    \caption{Ellipse parameters over time for voltage and current curves in the experimental setup, with a resistor plus inductor fault.}
    \label{fig:FLL_Outkontactor}
\end{figure}
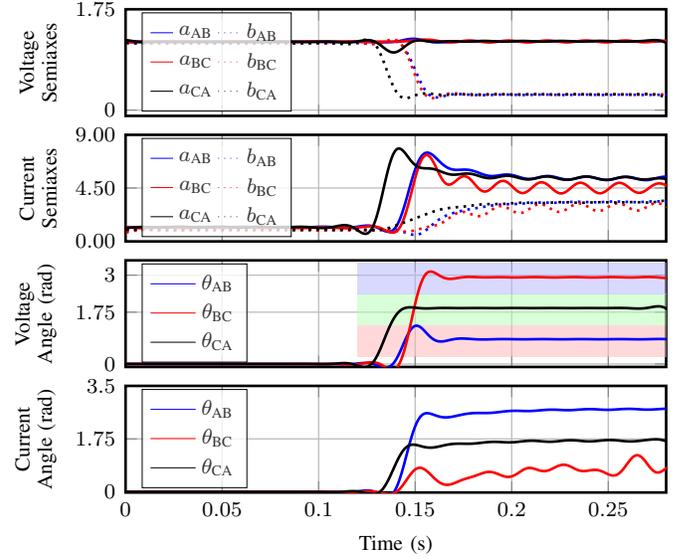

Fig. \ref{fig:FLL_Outkontactor} shows the results of the analysis for each type of LLF. Based on the voltage behavior, fault classification can be conducted with the help of Fig. \ref{fig:fault_classification_circle}. The correspondence between the experimental and simulated results, as shown in Fig. \ref{fig:FLL_Pex_RL} is clear for both voltage and current characteristics. Therefore, the proposed method is consistent for the study and characterisation of electrical faults, demonstrating its efficacy and utility for this task.

\begin{figure}[]
    \centering
    \begin{tikzpicture}
        \begin{axis}[
            width=0.99\columnwidth,
            height=3cm,
            xmin=0, xmax=0.28,
            ymin=-1.6, ymax=1.6,
            ylabel={{Voltage \\ (pu)}},
            ylabel style={font=\footnotesize, align=center},
            grid=major,
            grid style={line width=0.5pt, gray!50},
            line width=1pt,
            xticklabel=\empty,
            xtick={0,0.05,0.1,0.15,0.2,0.25,0.3,0.35,0.4,0.45},
            ytick={-1,0,1},
            yticklabel={\num[round-mode=places, round-precision=2]{\tick}},
            tick label style={font=\footnotesize},
            legend pos=north west,
            legend style={font=\footnotesize, fill=white, fill opacity=0.8, line width=0.3pt},
            name=voltage_plot,
            enlarge x limits=false,
            enlarge y limits=false,
        ]
            \addplot[red, solid] table[x=t, y=va, col sep=comma] {data/Falla_A-B-C.csv};
            \addplot[green!70!black, solid] table[x=t, y=vb, col sep=comma] {data/Falla_A-B-C.csv};
            \addplot[blue, solid] table[x=t, y=vc, col sep=comma] {data/Falla_A-B-C.csv};
            
            \legend{$v_A$, $v_B$, $v_C$}
        \end{axis}
        
        \begin{axis}[
            width=0.99\columnwidth,
            height=3cm,
            xmin=0, xmax=0.28,
            ymin=-0.3, ymax=0.3,
            xlabel style={font=\footnotesize},
            ylabel={Current\\(pu)},
            ylabel style={font=\footnotesize, align=center},
            grid=major,
            grid style={line width=0.5pt, gray!50},
            line width=1pt,
            at={(voltage_plot.below south west)},
            anchor=north west,
            yshift=0cm,
            xticklabel=\empty,
            xtick={0,0.05,0.1,0.15,0.2,0.25,0.3,0.35,0.4,0.45},
            ytick={-0.2,0,0.2},
            yticklabel={\num[round-mode=places, round-precision=2]{\tick}},
            tick label style={font=\footnotesize},
            legend pos=north west,
            legend style={font=\footnotesize, fill=white, fill opacity=0.8, line width=0.3pt},
            name=current_plot,
            enlarge x limits=false,
            enlarge y limits=false,
        ]
            \addplot[red, solid] table[x=t, y=ia, col sep=comma]{data/Falla_A-B-C.csv};
            \addplot[green!70!black, solid] table[x=t, y=ib, col sep=comma]{data/Falla_A-B-C.csv};
            \addplot[blue, solid] table[x=t, y=ic, col sep=comma] {data/Falla_A-B-C.csv};
            
            \legend{$i_A$, $i_B$, $i_C$}
        \end{axis}

        \begin{axis}[
            width=0.99\columnwidth,
            height=3cm,
            xmin=0, xmax=0.28,
            ymin=-0.1, ymax=2.1,
            ylabel={Voltage\\ semiaxes},
            ylabel style={font=\footnotesize, align=center},
            grid=major,
            grid style={line width=0.5pt, gray!50},
            line width=1pt,
            at={(current_plot.below south west)},
            anchor=north west,
            yshift=0cm,
            xticklabel=\empty,
            xtick={0,0.05,0.1,0.15,0.2,0.25,0.3,0.35,0.4,0.45},
            ytick={0,1,2},
            yticklabel={$\phantom{-}$\num[round-mode=places, round-precision=2]{\tick}},
            tick label style={font=\footnotesize},
            legend pos=north west,
            legend style={font=\footnotesize, fill=white, fill opacity=0.8, line width=0.3pt},
            name=voltage_axes_plot,
            enlarge x limits=false,
            enlarge y limits=false,
        ]
            \addplot[cyan!50!blue, solid] table[x=t, y=R_v, col sep=comma] {data/Falla_A-B-C_ellip.csv};
            \addplot[orange!50!red, solid] table[x=t, y=r_v, col sep=comma] {data/Falla_A-B-C_ellip.csv};
            
            \legend{$a$, $b$}
        \end{axis}
        
        \begin{axis}[
            width=0.99\columnwidth,
            height=3cm,
            xmin=0, xmax=0.28,
            ymin=0, ymax=12,
            xlabel={Time (s)},
            xlabel style={font=\footnotesize},
            ylabel={Current\\semiaxes},
            ylabel style={font=\footnotesize, align=center},
            grid=major,
            grid style={line width=0.5pt, gray!50},
            line width=1pt,
            at={(voltage_axes_plot.below south west)},
            anchor=north west,
            yshift=0cm,
            xtick={0,0.05,0.1,0.15,0.2,0.25,0.3,0.35,0.4,0.45},
            xticklabel style={/pgf/number format/fixed},
            ytick={0,6,12},
            yticklabel={\num[round-mode=places, round-precision=2]{\tick}},
            tick label style={font=\footnotesize},
            legend pos=north west,
            legend style={font=\footnotesize, fill=white, fill opacity=0.8, line width=0.3pt},
            enlarge x limits=false,
            enlarge y limits=false,
        ]
            \addplot[cyan!50!blue, solid] table[x=t, y=R_i, col sep=comma] {data/Falla_A-B-C_ellip.csv};
            \addplot[orange!50!red, solid] table[x=t, y=r_i, col sep=comma] {data/Falla_A-B-C_ellip.csv};
            
            \legend{$a$, $b$}
        \end{axis}

    \end{tikzpicture}
    \caption{Three-phase voltage and current waveforms during A-B-C fault event.}
    \label{fig:F_ABC}
\end{figure}
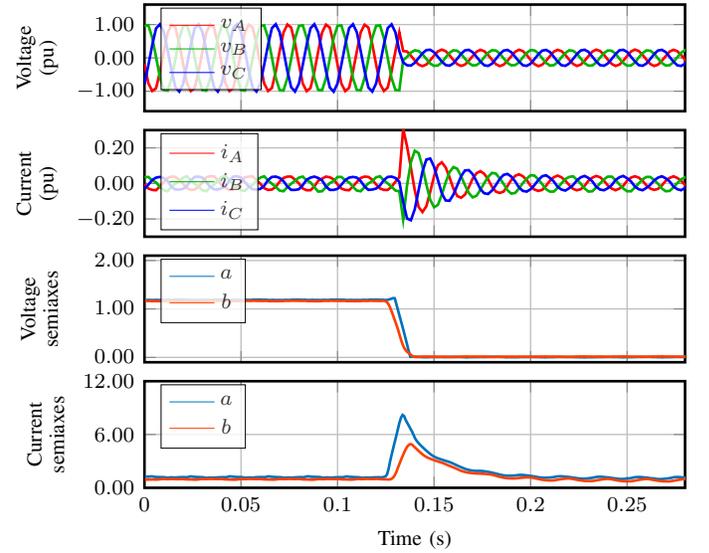

Finally, the proposed methodology has been applied to a symmetrical fault. Similar to line-to-line faults, the voltage and current curves reside within the Kirchhoff plane. Furthermore, the shape of the curve consistently resembles a circle, rendering the discussion of an inclination angle irrelevant; it may be asserted that this angle is zero. In Fig. \ref{fig:F_ABC}, both semi-axes remain constant throughout the duration of the fault, with the exception of a slight difference at the beginning, closely paralleling the behaviour of the current. This type of fault can be identified through a comparison of semi-axes.

\section{Conclusions}\label{sec:conclusions}

This paper has presented a novel approach for electrical fault detection and classification based on robust ellipse fitting techniques using Geometric Algebra. The proposed methodology offers a direct geometric characterization of voltage and current trajectories that captures the complete waveform evolution, addressing the significant challenges that conventional protection methods face in power systems with high penetration of inverter-based resources (IBRs).

The use of bivector plane orientation and its normalized components enables effective classification of line-to-ground faults, while ellipse parameters i.e. semi-axes and inclination angle, complement this classification to identify line-to-line and three-phase faults that are particularly problematic for traditional distance protection in IBR-dominated systems. The technique significantly reduces the necessary analysis window to just a quarter of a cycle, enabling faster fault detection while maintaining robustness against noise and transients. The method also handles degenerate cases where line-to-line faults produce straight-line trajectories of resistive fault or fault impedance close to zero.

Validation through simulation studies with IBR-based distribution grids and laboratory experiments with physical converters demonstrates that the proposed method effectively handles complex asymmetric faults in realistic scenarios with fitting errors consistently below 1\% for quarter-cycle windows under moderate noise conditions. The implemented geometric rotation overcomes the fundamental limitations of the conventional Clarke transform by preserving the original three-dimensional shape of voltage curves, particularly when zero-sequence components are present during unbalanced conditions.

The method provides quantitative fault classification capabilities for power system monitoring and protection in modern grids, offering a promising solution to the protection challenges introduced by the increasing integration of renewable energy sources through power electronic converters.

\begin{appendices}
\section{Geometric Algebra for Power System Analysis}
\renewcommand{\thesubsection}{\Alph{subsection}}

\subsection{Algebraic Structure and Multivector Space}

Given an orthonormal basis $\bm{\sigma}=\{\bm{\sigma}_1,\bm{\sigma}_2,\ldots,\bm{\sigma}_n\}$ for the vector space $\mathbb{R}^n$, we construct the Euclidean geometric algebra $\mathcal{G}_n$ with signature $(n,0,0)$. The basis elements satisfy the fundamental relations:

\begin{equation}
\bm{\sigma}_i^2 = 1, \quad \bm{\sigma}_i \bm{\sigma}_j = -\bm{\sigma}_j \bm{\sigma}_i \quad \text{for } i \neq j
\end{equation}

For the specific case of $n=3$, the complete basis for $\mathcal{G}_3$ consists of $2^3 = 8$ elements:
\begin{equation}
\{1, \bm{\sigma}_1, \bm{\sigma}_2, \bm{\sigma}_3, \bm{\sigma}_{12}, \bm{\sigma}_{13}, \bm{\sigma}_{23}, \bm{\sigma}_{123}\}
\end{equation}

where $\bm{\sigma}_{ij} = \bm{\sigma}_i \wedge \bm{\sigma}_j = \bm{\sigma}_i \bm{\sigma}_j$ for $i \neq j$, and $\bm{\sigma}_{123} = \bm{\sigma}_1 \wedge \bm{\sigma}_2 \wedge \bm{\sigma}_3$.

The wedge product $\wedge$ is the antisymmetric exterior or Grassmann  product that constructs higher-grade multivectors from lower-grade ones, representing oriented geometric objects such as planes (bivectors) and volumes (trivectors). For vectors, it satisfies $\bm{a} \wedge \bm{b} = -\bm{b} \wedge \bm{a}$ and vanishes when the operands are parallel.

A general multivector $\bm{M} \in \mathcal{G}_3$ can be expressed as:
\begin{equation}
\bm{M} = \sum_{k=0}^{3} \langle\bm{M}\rangle_k
\end{equation}

where $\langle\bm{M}\rangle_k$ denotes the $k$-vector part (grade-$k$ component) of $\bm{M}$.

\subsection{Bivector Algebra and Plane Representation}

Bivectors in $\mathcal{G}_3$ form a three-dimensional subspace spanned by $\{\bm{\sigma}_{12}, \bm{\sigma}_{13}, \bm{\sigma}_{23}\}$. A general bivector $\bm{B}$ can be written as:

\begin{equation}
\bm{B} = B_{12}\bm{\sigma}_{12} + B_{13}\bm{\sigma}_{13} + B_{23}\bm{\sigma}_{23}
\end{equation}

The fundamental properties of bivectors include:

\begin{align}
\bm{\sigma}_{ij}^2 &= -1 \\
\bm{\sigma}_{ij}\bm{\sigma}_{jk} &= \bm{\sigma}_{ik} 
\end{align}

The magnitude is computed as:
\begin{equation}
\|\bm{B}\| = \sqrt{B_{12}^2 + B_{13}^2 + B_{23}^2}
\end{equation}

thus, a unit bivector is $\hat{\bm{B}} = \bm{B}/\|\bm{B}\|$.
\subsection{Geometric Product and Contraction Operations}

The geometric product between bivectors $\bm{A}$ and $\bm{B}$ decomposes into grade-separated components:

\begin{equation}
\begin{aligned}
\bm{A}\bm{B} &= \bm{A} \rfloor \bm{B} + \bm{A} \times \bm{B} +\bm{A} \wedge \bm{B} \\
&=\langle\bm{A}\bm{B}\rangle_0 + \langle\bm{A}\bm{B}\rangle_2 + \langle\bm{A}\bm{B}\rangle_4
\end{aligned}
\end{equation}

The scalar part (contraction) is given by:
\begin{equation}
\bm{A} \rfloor \bm{B}  = \langle\bm{A}\bm{B}\rangle_0=  \frac{1}{2}(\bm{A}\bm{B} + \bm{B}\bm{A}) - \bm{A} \wedge \bm{B}
\end{equation}

The grade two element is given by:
\begin{equation}
\bm{A} \times \bm{B} = \langle\bm{A}\bm{B}\rangle_2= \frac{1}{2}(\bm{A}\bm{B} - \bm{B}\bm{A}) 
\end{equation}
Note that the $\times$ symbol represents the \textit{commutator} product in geometric algebra, similar to the commutator in Lie Algebra, but distinct from the conventional vector cross product.
The grade four element is given by:
\begin{equation}
\bm{A} \wedge \bm{B} = \langle\bm{A}\bm{B}\rangle_4
\end{equation}

\subsection{Rotor Algebra and Rotational Transformations}

A rotor $\bm{R}$ in $\mathcal{G}_3$ is an even multivector satisfying $\bm{R}\bm{R}^{\dagger} = 1$, where $\dagger$ denotes reversion:

\begin{equation}
\bm{R}^{\dagger} = \langle\bm{R}\rangle_0 - \langle\bm{R}\rangle_2
\end{equation}

The general form of a rotor in $\mathcal{G}_3$ is:
\begin{equation}
\bm{R} = \cos\frac{\theta}{2} + \sin\frac{\theta}{2}\hat{\bm{L}}
\end{equation}

where $\theta$ is the rotation angle and $\hat{\bm{L}}$ is a unit bivector representing the rotation plane.

For bivector-to-bivector rotations, given unit bivectors $\hat{\bm{A}}$ and $\hat{\bm{B}}$, the rotor that rotates $\hat{\bm{B}}$ to $\hat{\bm{A}}$ is:

\begin{equation}
\bm{R} = \frac{1 + \hat{\bm{A}}\hat{\bm{B}}^{\dagger}}{\|1 + \hat{\bm{A}}\hat{\bm{B}}^{\dagger}\|}
\label{eq:Rotations_components}
\end{equation}

provided $\hat{\bm{A}} \neq -\hat{\bm{B}}$.

\subsection{Invariants and Geometric Measures}

The angle between two bivectors $\bm{A}$ and $\bm{B}$ is determined by:
\begin{equation}
\cos\theta = \frac{\bm{A} \rfloor \bm{B}^{\dagger}}{\|\bm{A}\|\|\bm{B}\|} = \hat{\bm{A}} \rfloor \hat{\bm{B}}^{\dagger}
\end{equation}

For the Kirchhoff plane bivector $\bm{K} = \bm{\sigma}_{12} + \bm{\sigma}_{23} + \bm{\sigma}_{31}$, the normalized form is:
\begin{equation}
\hat{\bm{K}} = \frac{1}{\sqrt{3}}(\bm{\sigma}_{12} + \bm{\sigma}_{23} + \bm{\sigma}_{31})
\end{equation}

The deviation of a system bivector $\bm{B}$ from the Kirchhoff plane provides a direct measure of three-phase unbalance:
\begin{equation}
\delta = \arccos\left(\frac{\bm{B} \rfloor \hat{\bm{K}}^{\dagger}}{\|\bm{B}\|}\right) = \arccos\left(\hat{\bm{B}} \rfloor \hat{\bm{K}}^{\dagger}\right) 
\end{equation}

\end{appendices}

\bibliographystyle{IEEEtran}
\bibliography{bib}

\end{document}